\documentclass[nohyper]{JHEP}

\usepackage{amsmath}
\usepackage{amscd}
\usepackage{amsfonts}
\usepackage{amsthm}
\usepackage{graphicx}
\allowdisplaybreaks[1]

\def\IR{\mathbb{R}}

\newcommand{\tr}{\mbox{${\rm tr \;}$}}

\newsavebox{\dottedsquare}
\newsavebox{\Fcirc}
\newsavebox{\sdot}
\savebox{\Fcirc}(0,0){
\put(0,0){\circle*{1.5}}
}
\savebox{\sdot}(0,0){
\put(0,0){\circle*{0.2}}
}

\savebox{\dottedsquare}(20,20)[bl]{
\multiput(0,0)(2,0){10}{\usebox{\sdot}}
\multiput(0,0)(0,2){10}{\usebox{\sdot}}
\multiput(20,0)(0,2){11}{\usebox{\sdot}}
\multiput(0,20)(2,0){11}{\usebox{\sdot}}
\multiput(0,0)(20,0){2}{\usebox{\Fcirc}}
\multiput(0,20)(20,0){2}{\usebox{\Fcirc}}
}

\newcommand{\rs}[1]{\raisebox{1.5ex}[0pt][0pt]{#1}}

\title{The non-abelian open superstring effective action through order $ \alpha'{}^3$}

\author{by Paul Koerber\thanks{Aspirant FWO} and Alexander
Sevrin\\
    Theoretische Natuurkunde, Vrije Universiteit Brussel \\
    Pleinlaan 2, B-1050 Brussels, Belgium\\
        E-mail: \email{koerber, asevrin@tena4.vub.ac.be}}

\preprint{VUB/TENA/01/05\\ \hepth{0108169}}

\abstract{Using the method developed in {\tt hep-th/0103015}, we determine
the non-abelian Born-Infeld action through $ {\cal O}(\alpha'{}^3)$.
We start from solutions to a Yang-Mills theory which define a
stable holomorphic vector bundle. In D-brane context this corresponds
to BPS configurations in the limit of small background fields. Subsequently
we investigate its deformation away from this limit. Through
$ {\cal O}(\alpha'{}^2)$, a unique, modulo field redefinitions, solution
emerges. At $ {\cal O}(\alpha'{}^3)$ we find a one-parameter family of allowed
deformations. The presence of derivative terms
turns out to be essential. Finally, we present a detailed comparison of our results
to existing, partial results.}


\begin{document}
\section{Introduction}

The bosonic massless degrees of freedom of an open string ending on a
flat D$p$-brane are a $U(1)$ gauge field, associated to excitations of
the string longitudinal to the brane, and neutral scalar fields,
describing the fluctuations of the brane in the transverse directions
\cite{pol}. For slowly
varying fields, i.e.\ ignoring derivative terms, the effective action for these
massless degrees of freedom is known through all orders in $\alpha'$. It is the
$d=10$ supersymmetric abelian Born-Infeld theory, dimensionally reduced to
$p+1$ dimensions \cite{BI}, \cite{susynbi}.

Once several, say $n$, D-branes coincide, the gauge group gets enhanced
from $U(1)$ to $U(n)$, \cite{witten}. In leading order in $\alpha'$, the
effective action is precisely the $d=10$ supersymmetric $U(n)$ Yang-Mills
theory dimensionally reduced to $p+1$ dimensions. The exact structure of the
full effective action remains an elusive puzzle. Two complications arise in
comparison with the abelian case:
\begin{itemize}
\item Because all fields transform in the adjoint of $U(n)$, an ordering
prescription is needed.
\item There is no covariant notion of a slowly varying field. In other words,
higher order derivative terms have to be included.
\end{itemize}

The calculation of open superstring amplitudes allows for a direct
determination of the effective action. This approach lead to firm results
through order $\alpha'{}^2$ \cite{direct}, \cite{direct1}, \cite{bilal}. 
A first systematic investigation of the effective action was performed in \cite{Tstr}.
A first observation is that, 
as the effective action has to match gluon disk amplitudes, there is necessarily
only a single overall group trace. Furthermore,
the effective action $ {\cal S}$ is necessarily of the form
${\cal S}={\cal S}_1+{\cal S}_2+{\cal S}_3$. Here, ${\cal S}_1$
does not contain any covariant derivatives acting on the
fieldstrength and is, by definition,
the non-abelian Born-Infeld action.
Both ${\cal S}_2$ and ${\cal S}_3$ contain the terms with derivatives acting on the
fieldstrength, but while ${\cal S}_2$ has only terms with symmetrized 
products of covariant derivatives, ${\cal S}_3$ has anti-symmetrized products
of covariant derivatives as well. Because of
$[D,D]\cdot=[F,\cdot ]$, the separation between ${\cal S}_1$ and ${\cal S}_3$
is not unambiguously defined. This ambiguity was fixed in \cite{Tstr} by proposing
that ${\cal S}_1$ is the non-abelian
Born-Infeld action defined by means of the symmetrized trace prescription.
It assumes the same form as the abelian Born-Infeld action but upon expanding the action
in powers of the fieldstrength, one first symmetrizes all terms and subsequently one performs
the group trace. Indeed, all other terms without derivatives 
not belonging to this class can be rewritten
as elements of ${\cal S}_3$.
In the abelian limit, ${\cal S}_1$ reduces then to the abelian Born-Infeld action, ${\cal S}_3$
vanishes and ${\cal S}_2$, which is present \cite{AT},
vanishes in the limit of slowly varying fields. Through order $\alpha'{}^2$,
${\cal S}_2$ and ${\cal S}_3$ vanish.  At higher orders, contributions to those terms are expected
as well. Indeed, the results in \cite{HT} and
\cite{DST} demonstrated that knowledge of ${\cal S}_1$ is not sufficient to reproduce even
simple features of D-brane dynamics. As we will show in this paper, from order 
$\alpha'{}^3$ on, ${\cal S}_2$ and ${\cal S}_3$ receive non-trivial contributions
as well.

A direct calculation at higher orders becomes technically very involved (however see \cite{kit} and \cite{bilal2}
for partial results in this direction), so one
is forced to develop alternative approaches.
One of these, motivated by the
results in \cite{HT} and \cite{DST}, uses the mass spectrum as a guideline
which resulted in partial higher order results through order $\alpha'{}^4$
\cite{STT}.

The problem at hand possesses a lot of supersymmetry: there are 16 linearly and
16 non-linearly realized supersymmetries. This rises the hope that requiring
the deformations of the Yang-Mills action to be supersymmetric will severely
restrict the possibilities, perhaps even leading to a unique all order result
\cite{jim}, \cite{BRS}.
Recently it was shown at component level that 
supersymmetry indeed almost fully determines the action
through $ {\cal O}(\alpha'{}^2)$ including fermionic and derivative terms
\cite{goteborg}. The presence of both linear and non-linear supersymmetry would
suggest the existence of an explicitly $\kappa$-invariant formulation. Despite
an attempt at lower order \cite{kappa}, this remains an open problem
\cite{bilal}.

A related approach was recently developed in \cite{milaan}.
Starting from the $N=4$, $d=4$ supersymmetric Yang-Mills theory, the bosonic part of the
one-loop effective action through operators of dimension 10 was calculated. If one
assumes that the supersymmetric deformation of the Yang-Mills action is unique,
then this calculation should yield the non-abelian open superstring action. However, as we will
discuss in section 6, this is not so. In addition, the method of \cite{milaan} is restricted 
to four dimensions.

A very different approach was launched in \cite{nieuw}. Starting point was the
existence of a particular class of solutions to Yang-Mills generalizing the
usual instantons in four dimensions. These solutions define stable holomorphic
bundles \cite{durham}. In the context of D-brane physics, such solutions
correspond to BPS solutions in the weak field limit \cite{GSW}. In
\cite{nieuw}, deformations of such solutions were analyzed in the abelian
limit. Arbitrary powers of the
fieldstrength were added to the Yang-Mills action. Subsequently it was required
that stable holomorphic bundles, or some deformation thereof, still provides
solutions to the equations of
motion. Surprisingly this approach leads to a {\em unique} deformation: the
abelian Born-Infeld action. While the holomorphicity condition remains
unchanged, the stability condition acquires higher order corrections as
well.

An obvious question which arises in this context is whether the method sketched
above leads to similar restrictions in the non-abelian case. The analysis in
\cite{nieuw} used the explicit assumption that only
slowly varying fields appeared. In other words, the Yang-Mills action was
deformed by adding arbitrary powers of the fieldstrength to it, but terms
containing derivatives of the fieldstrength were excluded. In the non-abelian
case, there is no covariant notion of acceleration terms.
Indeed, in the abelian case, it is not hard to find a rescaling of the
coordinates and the gauge fields such that a limit exists where
the fieldstrength remains invariant
but its derivatives vanish. Such a limit does not exist in the non-abelian
case.
However, one might still try to repeat the analysis of \cite{nieuw} under the
same assumptions.
When doing this, we found no contribution at $ {\cal O}(\alpha'{}^3)$,
but we encountered an obstruction at order $\alpha'{}^4$
(i.e.\ order $F^6$).
This clearly showed the need to include derivative terms as well in the non-abelian
case. For alternative arguments we refer to the introduction of
\cite{bilal2} (see also \cite{kit} and \cite{milaan}).

As a first test, we analyze the deformation of the non-abelian Yang-Mills
action through order $\alpha'{}^3$. At this order partial results were obtained
before \cite{kit}, \cite{bilal2}, \cite{milaan}. A clear full answer is however still
lacking. The present method
shows a major drawback once derivative terms are allowed: the number of terms
which can potentially contribute to the action increases dramatically with each
order in $\alpha'{}$. An additional difficulty is that, because of partial
integration, Bianchi identities and the $[D,D]\cdot=[F,\cdot]$ identity, many
relations between various terms exist. In order to deal with this in an
efficient way, we wrote a program \cite{paul} in Java,
an object oriented language based on the syntax of C,
 which classifies at a given order in
$ \alpha'$ the independent terms in the action, calculates the resulting
equations of motion and analyzes the deformations of the stability condition.
Finally it takes care of field redefinitions as well.

This paper is organized as follows. In the next section we briefly review the
relevant solutions in Yang-Mills. In sections 3 to 5 we study the deformations
through order $ \alpha'{}^3$ and systematize our method. In the last section we analyze our results and
confront them with previously known results. The first appendix explains our
conventions and notations.  Appendix \ref{order2} gives the equations of motions at order $ \alpha'{}^2$ and
appendix \ref{lagduy3} a base for the lagrangian and the stability condition deformation at order $\alpha'{}^3$.

\section{The leading order: stable holomorphic bundles in Yang-Mills}
\label{leading}

In leading order, the effective action of $n$ coinciding D$p$-branes is the
supersymmetric $U(n)$ Yang-Mills theory in ten
dimensions\footnote{We ignore an
overall multiplicative constant.},
\begin{eqnarray}
{\cal S}=\int d^{10}x\,\tr\left(-\frac 1 4 F_{\mu \nu}F^{\mu \nu}+\frac i
2\bar\psi D\!\!\!\!/\psi\right), \label{ac1}
\end{eqnarray}
dimensionally reduced to $p+1$ dimensions.
The Majorana-Weyl spinor $\psi$ transforms in
the adjoint representation of $U(n)$. In this way we get as world-volume
degrees of freedom a $U(n)$ gauge field in $p+1$ dimensions, $9-p$ scalar
fields and the 16 components of $\psi$.

In the present paper, we will ignore the transversal scalars and the fermionic
degrees of freedom as they do not seem to give additional information.
Furthermore, we make one important assumption: instead of restricting ourselves
to $d=10$ or less, we will require our analysis to hold in any
even dimension! In
this way we avoid relations which exist in particular dimensions.

Our starting point is a $U(n)$ Yang-Mills action in an even dimensional, flat
Euclidean space\footnote{As the metric is +1 in all directions, we simplify the
notation by putting all indices down. Unless stated otherwise, we
sum over repeated indices. Furthermore, the lagrangian eq.\ (\ref{lag0}), should still
be multiplied by an arbitrary coupling constant $-1/g^2$.},
\begin{eqnarray}
{\cal L}_{(0)}=\frac{1}{4}\tr
F_{\mu_1\mu_2}F_{\mu_2\mu_1}.\label{lag0}
\end{eqnarray}
In complex coordinates, the equations of motion, $D_\nu F_{\nu\mu}=0$, read,
\begin{eqnarray}
0&=&D_{\bar\alpha }F_{\alpha \bar\beta}+D_{\alpha }F_{\bar\alpha
\bar\beta}\nonumber\\
&=&D_{\bar\beta}F_{\alpha \bar\alpha }+2D_{\alpha }F_{\bar\alpha
\bar\beta}, \label{eom0}
\end{eqnarray}
where we used the Bianchi identities in the last line. One sees that
configurations satisfying
\begin{eqnarray}
F_{\alpha\beta}=F_{\bar\alpha\bar\beta}=0, \label{hol}
\end{eqnarray}
and
\begin{eqnarray}
g^{\alpha\bar\beta}F_{\alpha\bar\beta}=\sum_{\alpha}F_{\alpha\bar\alpha}
\equiv F_{\alpha\bar \alpha}=0,\label{duy}
\end{eqnarray}
solve the equations of motion \cite{durham}. Eq.\ (\ref{hol})
defines a holomorphic bundle, while eq.\ (\ref{duy}) guarantees the
stability of the bundle \cite{DUY}.  We will alternatively call the latter equation the
Donaldson-Uhlenbeck-Yau condition, henceforth abbreviated DUY condition.

Restricting to dimensions less than 10, these solutions are BPS configurations
of D-branes in the limit where $2\pi\alpha{}'F$ is small.
When $p=2$,  the BPS conditions are recognized as the standard instanton
equations. Note that constant magnetic background fields which satisfy the
conditions eqs.\ (\ref{hol}) and (\ref{duy}) can be reinterpreted, after
T-dualization, as BPS configurations of D$p$-branes at angles \cite{Angles}.

In the next sections, we will investigate order by order in $\alpha'$ the most
general deformation of eq.\ (\ref{lag0}). So we will add at each order the most
general polynomial in the fieldstrength and its covariant derivatives, each
term with an arbitrary coupling constant. Subsequently we will demand that
configurations of the form eqs.\ (\ref{hol}) and (\ref{duy}) solve the deformed
equations of motion. As it turns out this will fix, modulo certain field
redefinitions, the coupling constants. Simultaneously, we will have to deform
the stability condition eq.\ (\ref{duy}) as well. Concerning the deformation of
eq.\ (\ref{duy}), we require it to be such that it fully determines
$F_{\alpha\bar\alpha}$ i.e.\ {\em it should be such that $F_{\alpha\bar\alpha}$
only appears at the left-hand side
of the equation}.

\section{Learning from the low orders}
\subsection{The $\alpha'{}^1$ corrections}
\label{order1}

The most general deformation of the Yang-Mills action, ${\cal L}_{(0)}$,
at the next order is given by
$ {\cal L}_{(0)}+{\cal L}_{(1)}$ with,
\begin{eqnarray}
{\cal L}_{(1)}=2\pi\alpha'l^3_{0,0,0} \tr \left(
F_{\mu_1\mu_2}F_{\mu_2\mu_3}F_{\mu_3\mu_1}\right) +
2\pi\alpha'l^3_{1,1,0}
\tr\left(\left(D_{\mu_3}D_{\mu_1}F_{\mu_1\mu_2}\right)
F_{\mu_2\mu_3}\right),\label{lag1}
\end{eqnarray}
with $l^3_{0,0,0}$ and $l^3_{1,1,0}$ arbitrary constants.
This expression takes into account partial
integration, Bianchi and $[D,D]\cdot=[F,\cdot]$ identities,
as we will study in more detail in section \ref{algorithm}.
In order to simplify our notation, we will put $2\pi\alpha'=1$
from now on.
In complex coordinates, one finds that the equations of motion following from
eq.\ (\ref{lag1}) read as
\begin{eqnarray}
0&=&D_{\bar{\beta}}F_{\alpha\bar\alpha}-3 \; l^3_{0,0,0}
\left(D_{\bar{\beta}}F_{\alpha_1\bar{\alpha}_2}\right)
F_{\alpha_2\bar{\alpha}_1}
+3 \; l^3_{0,0,0}
F_{\alpha_1\bar{\alpha}_2}\left(D_{\bar{\beta}}F_{\alpha_2\bar{\alpha}_1}
\right)
+\nonumber\\
&&2 \; l^3_{1,1,0}
\left(D_{\bar{\beta}}F_{\alpha_1\bar{\alpha}_1}\right)F_{\alpha_2
\bar{\alpha}_2}
-2 \; l^3_{1,1,0}
F_{\alpha_1\bar{\alpha}_1}\left(D_{\bar{\beta}}F_{\alpha_2\bar{\alpha}_2}
\right)
+\nonumber\\ &&\left(4 \; l^3_{1,1,0}+3 \; l^3_{0,0,0}\right)
\left(D_{\bar{\beta}}D_{\bar{\alpha}_1}D_{\alpha_1}F_{\alpha_2\bar{\alpha}_2}
\right)
-3 \; l^3_{0,0,0}
\left(D_{\alpha_1}D_{\bar{\beta}}D_{\bar{\alpha}_1}F_{\alpha_2\bar{\alpha}_2}
\right) ,\label{eom1}
\end{eqnarray}
where we used eq.\ (\ref{hol}) and the Bianchi identities. Almost all terms
vanish when implementing eq.\ (\ref{duy}), leaving only the second and third
term.
At this order the only allowed deformation\footnote{We remind the reader that
we view the deformed stability condition as an expression for
$F_{\alpha\bar\alpha}$. So while additional deformation terms of the form
$F_{\alpha_1\bar{\alpha}_1}F_{\alpha_2\bar{\alpha}_2}$ or
$\left(D_{\bar{\alpha}_1}D_{\alpha_1}F_{\alpha_2\bar{\alpha}_2}\right)$ are
dimensionally allowed, they are excluded by our ansatz.} of eq.\
(\ref{duy}) is,
\begin{eqnarray}
0=F_{\alpha\bar\alpha}+ d^3_{0,0,0}
F_{\alpha_1\bar{\alpha}_2}F_{\alpha_2\bar{\alpha}_1}.
\end{eqnarray}
One sees immediately that the equations of motion are solved provided we do not
deform eq.\ (\ref{duy}), i.e.\ $d^3_{0,0,0}=0$ and we put $l_{0,0,0}^3=0$. This
eliminates the second and the third term in eq.\ (\ref{eom1}). The remainder of
the equation of motion is satisfied by virtue of eq.\ (\ref{duy}).
The surviving second
term  in $ {\cal L}_{(1)}$ can then be eliminated by a field redefinition,
\begin{eqnarray}
A_\mu\longrightarrow A_\mu - l^3_{1,1,0} D_\nu F_{\nu\mu},
\end{eqnarray}
which exhausts the field redefinitions at this order. Concluding the $ {\cal
O}( \alpha')$
deformation of both the Yang-Mills action and the stability condition
vanish, which is consistent with direct calculations.

\subsection{The $\alpha'{}^2$ corrections}

At the next order the most general deformation of the Yang-Mills lagrangian
reads as
\begin{eqnarray}
{\cal L}= {\cal L}_{(0)}+{\cal L}_{(2)},
\end{eqnarray}
where $ {\cal L}_{(0)}$ is given in eq.\ (\ref{lag0}) and $ {\cal L}_{(2)}$ is
\begin{eqnarray}
{\cal L}_{(2)}&=&\tr\left(
l^4_{0,0,0} F_{\mu_1\mu_2}F_{\mu_2\mu_3}F_{\mu_3\mu_4}F_{\mu_4\mu_1} +
l^4_{ 0,0,1}F_{\mu_1\mu_2}F_{\mu_2\mu_3}F_{\mu_4\mu_1}F_{\mu_3\mu_4} +
\right. \nonumber \\
&&l^4_{0,1,0}F_{\mu_1\mu_2}F_{\mu_2\mu_1}F_{\mu_3\mu_4}F_{\mu_4\mu_3}+
l^4_{0,1,1}F_{\mu_1\mu_2}F_{\mu_3\mu_4}F_{\mu_2\mu_1}F_{\mu_4\mu_3} +
\nonumber \\
&&l^4_{1,2,1}
\left(D_{\mu_4}D_{\mu_1}F_{\mu_1\mu_2}\right)F_{\mu_3\mu_4}F_{\mu_2\mu_3} +
l^4_{1,2,6}\left(D_{\mu_1}F_{\mu_1\mu_2}\right)F_{\mu_2\mu_3}\left(D_{\mu_4}F_{
\mu_3\mu_4}\right) +\nonumber \\
&&\left. l^4_{2,1,23}
\left(D_{\mu_4}D_{\mu_1}F_{\mu_1\mu_2}\right)\left(D_{\mu_4}D_{\mu_3}
F_{\mu_2\mu_3}\right)\right).
\label{lag2}\end{eqnarray}
This is the most general deformation at order $ \alpha'{}^2$ where we used the
Bianchi, partial integration and $[D,D]\cdot =[F,\cdot ]$
identities. Both the deformation of the stability condition, eq.\ (\ref{duy}) and
the contribution to the equations of motion at this order are explicitly given
in appendix \ref{order2}.

As an illustration, we analyze the equations of motion in
some detail. It is clear
that the equations of the type $e^4_{0,0,s_3}$ and $e^4_{0,3,s_3}$ are
satisfied provided
\begin{eqnarray}
&&d^4_{0,0,0}=d^4_{0,0,1}=4l^4_{0,0,0}=2l^4_{0,0,1}=
-8l^4_{0,1,0}=-16l^4_{0,1,1}, \nonumber\\&&l^4_{1,2,1}=0.\label{res4}
\end{eqnarray}
The contributions of the type $e^4_{1,0,s_3}$ vanish, provided
\begin{eqnarray}
d^4_{1,0,4}=d^4_{1,0,5}=0,
\end{eqnarray}
holds. The remainder of the equations of motion now trivially vanishes when
applying eq.\ (\ref{duy})\footnote{Of course these terms will contribute at
order $ \alpha'{}^4$ as a consequence of the deformation eq.\ (\ref{duy2}).
See section \ref{algorithm}, step 7.b and 7.c for a systematic approach.}.

We can fix one more parameter. Initially we had two choices: we could choose
the overall multiplicative constant in front of the action and we can fix the
scale of the fieldstrength $F$. These two arbitrary constants are fixed by
choosing the conventional factor $1/4$ in front of the leading term in the
action and by putting\footnote{The unconventional $-$ sign is due
to the fact that we choose an anti-hermitean basis for $u(n)$.} $l^4_{0,0,0}=-1/24$.

Having done this we fixed the deformation of the action completely modulo the
coupling constants $l^4_{1,2,6}$ and $l^4_{2,1,23}$.
However, we still have to consider field redefinitions. The most general
field redefinition relevant at this order is
\begin{eqnarray}
A_\nu&\rightarrow& A_\nu +
f^3_{0,1,0} \left(D_{\mu_1}F_{\mu_1\mu_2}\right)F_{\mu_2\nu}+
f^3_{0,1,1} F_{\mu_2\nu}\left(D_{\mu_1}F_{\mu_1\mu_2}\right)+ \nonumber\\
&&f^3_{0,1,2} F_{\mu_1\mu_2}\left(D_{\mu_1}F_{\mu_2\nu}\right)+
f^3_{0,1,3} \left(D_{\mu_1}F_{\mu_2\nu}\right)F_{\mu_1\mu_2} + \nonumber\\
&&f^3_{1,0,1} \left(D_{\mu_1}D_{\mu_2}D_{\mu_1}F_{\mu_2\nu}\right),
\end{eqnarray}
where again we took Bianchi identities and the $[D,D]\cdot=[F,\cdot]$ relation
into account. The coupling constants transform as follows under the
field redefinitions,
\begin{eqnarray}
l^4_{0,s_2,s_3}&\rightarrow& l^4_{0,s_2,s_3}, \nonumber\\
l^4_{1,2,1}&\rightarrow &l^4_{1,2,1}-f^3_{0,1,2}+2f^3_{1,0,1}+
f^3_{0,1,3}, \nonumber\\
l^4_{1,2,6}&\rightarrow &l^4_{1,2,6} +f^3_{0,1,1}+
f^3_{0,1,2}-f^3_{0,1,0}-f^3_{0,1,3}, \nonumber\\
l^4_{2,1,23}&\rightarrow &l^4_{2,1,23}+f^3_{1,0,1}.
\end{eqnarray}
Taking into account that we have to keep $l^4_{1,2,1}=0$, we find that the
three field redefinitions
\begin{eqnarray}
f^3_{0,1,1}-f^3_{0,1,0}&=& -l^4_{1,2,6}+2 l^4_{2,1,23},\nonumber\\
f^3_{0,1,2}-f^3_{0,1,3}&=& -2l^4_{2,1,23},\nonumber\\
f^3_{1,0,1}&=&-l^4_{2,1,23} ,
\end{eqnarray}
eliminate the derivative terms in eq.\ (\ref{lag2}). This leaves us
two field redefinitions,
\begin{eqnarray}
A_\nu&\rightarrow& A_\nu +
\frac 1 2 \left(f^3_{0,1,0}+f^3_{0,1,1}\right)
\left\{ \left(D_{\mu_1}
F_{\mu_1\mu_2}\right),F_{\mu_2\nu}\right\}+
\nonumber\\
&&\frac 1 2 \left(f^3_{0,1,2}+ f^3_{0,1,3} \right)
\left\{F_{\mu_1\mu_2},\left(D_{\mu_1}F_{\mu_2\nu}\right)\right\},
\end{eqnarray}
which will not play any further role in this paper as they
only become potentially relevant at order $ \alpha'{}^4$ in the action.
Note that it is quite remarkable that certain terms which are removable
through field redefinitions, the $l^4_{1,2,1}$ term in casu, can get fixed
by our method.

Summarizing, through order $ \alpha'{}^2$, the lagrangian is given by
${\cal L}= {\cal L}_{(0)}+{\cal L}_{(2)}$, with ${\cal L}_{(0)}$ given in eq.\
(\ref{lag0}) and
\begin{eqnarray}
{\cal L}_{(2)}&=&-\tr\left(
\frac{1}{24} F_{\mu_1\mu_2}F_{\mu_2\mu_3}F_{\mu_3\mu_4}F_{\mu_4\mu_1} +
\frac{1}{12}F_{\mu_1\mu_2}F_{\mu_2\mu_3}F_{\mu_4\mu_1}F_{\mu_3\mu_4}-
\right. \nonumber \\
&& \left.\frac{1}{48}F_{\mu_1\mu_2}F_{\mu_2\mu_1}F_{\mu_3\mu_4}F_{\mu_4\mu_3}-
\frac{1}{96}F_{\mu_1\mu_2}F_{\mu_3\mu_4}F_{\mu_2\mu_1}F_{\mu_4\mu_3}\right).
\label{lag2def}
\end{eqnarray}
Configurations satisfying eq.\ (\ref{hol}) and
\begin{eqnarray}
0=F_{\alpha\bar\alpha}-
\frac 1 6 F_{\alpha_1\bar{\alpha}_2}F_{\alpha_2\bar{\alpha}_3}
F_{\alpha_3\bar{\alpha}_1} -
\frac 1 6 F_{\alpha_1\bar{\alpha}_2}F_{\alpha_3\bar{\alpha}_1}
F_{\alpha_2\bar{\alpha}_3},\label{duy2def}
\end{eqnarray}
solve the equations of motion. This result is again fully consistent with
direct calculations.

\section{Systematizing our method}
\label{algorithm}

After studying the low order cases, we are ready to put together the
calculational scheme.  Although this scheme can be slavishly followed
at higher orders, the calculations itself will become extremely lengthy.
So they were carried out by a computer program, written
specially for the task at hand. The language of choice was Java,
which as a modern object oriented programming language proved to
us more user-friendly than the in physics more commonly used C or
Fortran. In this section we will discuss what it does, and leave
the implementational details for what they are. More details as well as the source code will be given
in \cite{paul}.

Roughly put, the program will construct the most general lagrangian and
deformation of the DUY condition at each order in $\alpha'$.
Subsequently, we impose that fieldstrength configurations satisfying eq.\ (\ref{hol})
and the generalized DUY condition solve the equations of motion. This
generates a set of equations, since
the coefficient of each {\em independent} term in the equations of motion has
to be zero.  From
these conditions we can fix the coefficients of the lagrangian {\em as well} as
the coefficients of the DUY
deformation.  Having said this, we can delve into the technical details.

The program distinguishes 4 kinds of terms at each order: their properties are
listed in table \ref{tproperties}.
For each of these types, the program will have to:
\begin{enumerate}
\item Calculate all possible terms, using as building blocks antisymmetric
fields $F$ and covariant derivatives.
These terms have a priori arbitrary coefficients, which are labelled according
to the classification
scheme of appendix \ref{Conventions}.
\item Calculate all possible identities between those terms: these are the
partial integration identities (only for
the lagrangian), the Bianchi identities and the identities of the type
$[D,D]\cdot=[F,\cdot]$.
\item Solve those (linear) identities and thus separate the linearly dependent
terms from the linearly
independent terms, forming a base.  The program proceeds by eliminating one
term out of each equation.
Of course there is still the freedom of choosing the term to eliminate and this
will often give rise to some priority
rule.

Sometimes a term will carry some coefficient information. Upon
elimination, this information will be transferred
to the other terms in the equation.  An example: suppose $T^1, T^2, \ldots T^n$
are terms in
an equation of motion, which reads:
\begin{equation}
\sum_{j=1}^n c_j T^j = 0 \, .
\end{equation}
Term $T^j$ is said to {\em carry} its coefficient $c_j$.  Now, if $T^i$ is
eliminated from the identity:
\begin{equation}
T^i = \sum_{j \neq i} d_j T^j \, ,
\end{equation}
it has to transfer his coefficient $c_i$ to the other terms and the
equation
of motion becomes:
\begin{equation}
\sum_{j \neq i} (c_j + d_j c_i) T^j =0 \, .
\end{equation}
\end{enumerate}
\TABLE{
\centering
\begin{minipage}{10cm}
\begin{tabular}{|c||c|c|c|c|}
\hline
& & Field & DUY & Equations \\
\rs{Properties} & \rs{Lagrangian} & redefinitions & deformation & of motion \\
\hline\hline
Group trace & & &  & \\
implied\footnote{So there is cyclic symmetry} & \rs{yes} & \rs{no} & \rs{no} &
\rs{no} \\
\hline
Free index & no & yes & no & yes \\
\hline
Complex & & & & \\
coordinates\footnote{See appendix \ref{Conventions}.} & \rs{no} & \rs{no} &
\rs{yes} & \rs{yes} \\
\hline
Type of& Bianchi & Bianchi & Bianchi & Bianchi \\
identities           & PI\footnote{PI: partial integration identities.},
DDF\footnote{DDF: identities of the type $[D,D]\cdot=[F,\cdot]$.}
 & DDF$^d$ & DDF$^d$ & DDF$^d$ \\
\hline
Used & ${\cal L}$ & ${\cal F}$ & ${\cal D}$ & ${\cal E}$ \\
Symbols\footnote{The curly type is used for denoting all terms at a certain
order, the small type
in the labelling of the individual terms (as explained in appendix
\ref{Conventions}).} & $l$ & $f$ & $d$ & $e$ \\
\hline
\end{tabular}
\end{minipage}
\caption{Properties of the different types of
terms.}
\label{tproperties}
}

Because of the arbitrariness in choosing a base, many lagrangians will in fact
be equivalent.
When comparing with the results in the literature \cite{milaan} \cite{kit}, we
will have
to express their terms in our base.  Also, as an extra complication, we have to
take account of the fact
that the coefficients of some terms in the lagrangian may change when applying
a field redefinition and allow for them to differ
\cite{direct1}.  We call the latter from now on {\em FR changeable terms}.

After these initial remarks we state the algorithm, which has
to be repeated order per order\footnote{We will call the order in $F$ $n$,
so the order in $\alpha'$ is $n-2$.} in $\alpha'$:
\begin{enumerate}
\item Construct all possible terms in the lagrangian and all identities among
those.
\item If we want to know which terms are FR changeable, we must put in quite a
lot of extra work.
To the solution for the lagrangian already found at lower orders, apply a field
redefinition:
\begin{equation}
A_{\mu} \rightarrow A_{\mu} + {\cal F}_{\mu} \, ,
\end{equation}
where ${\cal F}_{\mu}$ is a linear combination of all possible {\em
independent} field
redefinition terms
of the appropriate lower orders.  Observe how the coefficients of the terms in
the lagrangian at the
present order change.
To clarify what we mean, we study the simplest case in detail.  At order
$\alpha'{}^0$,
the most general field redefinition is:
\begin{equation}
{\cal F}_{(0),\nu} = f^2_{0,0,0} D_{\mu_1} F_{\mu_1\nu} \, ,
\end{equation}
and, as we have already used in section \ref{order1}, the coefficient change of
term $l^3_{1,1,0}$ (see \eqref{lag1}) becomes
\begin{equation}
\Delta l^3_{1,1,0} = f^2_{0,0,0} \, .
\end{equation}
For each FR changeable term, we say that the term {\em carries}\footnote{An alternative
way of looking at this, is that the field redefinition terms form a dual vector space, because
the choice of a lagrangian term and a field redefinition term produces a (fractional) number, i.e.\
the coefficient change.  More precisely, the field redefinition terms are only a subspace of the dual vector space
and we want to make this clear by an appropriate choice of base of the original space.} his coefficient
change.  The other terms,
we will call {\em empty} i.e.\ carrying nothing.
\item Calculate all independent terms in the lagrangian.  When eliminating an
FR changeable term out of a
certain equation, its coefficient change must be transferred as explained,
because
we want to know how the remaining terms transform under a field redefinition.
So
all the other terms in that equation become FR changeable.
To minimize the amount of FR changeable terms we end up with, we use the
following priority rule:
eliminate empty terms first.  If we did not follow this
rule the FR changeability would quickly spread among all independent terms,
although many coefficient changes would be dependent\footnote{Note that our
``rule of thumb''
doesn't guarantee that {\em all} remaining coefficient changes are independent,
but certainly most
of them.}.

In the end, when comparing the result to the literature, we will only have to
consider empty terms.
\item Construct all possible {\em independent} field redefinition terms for
later use at higher orders.
\item Construct all possible {\em independent} contributions to the DUY
condition.  As an example, ${\cal D}_{(2)}$ from eq.\ \eqref{duy2} would be the
result at order $\alpha'{}^2$.
\item Construct all possible terms in the equations of motion.
\item The coefficients of those terms in the equations of motions will have
three contributions:
\begin{enumerate}
\item The coefficients obtained from varying the terms in the lagrangian
containing
the arbitrary lagrangian coefficients $l^n_{s_1,s_2,s_3}$.  Note that in the
non-abelian
case there is also a contribution from varying the covariant derivatives.
\item Subtraction of the deformed DUY condition:
\begin{equation}
D_{\bar{\beta}} ( F_{\alpha_1\bar{\alpha_1}} + \cdots + {\cal D}_{(n-3)} +
{\cal D}_{(n-2)}) = 0 \, .
\end{equation}
The first term cancels the contribution of the lagrangian to the equations of
motion at order
$\alpha'{}^0$.  The last term ${\cal D}_{(n-2)}$ contributes to the equations of
motion at the
present order $\alpha'{}^{(n-2)}$.
\item Consider a term with one or more 1-loops:
\begin{equation}
F_{\alpha_1\bar{\alpha}_1} \ldots F_{\alpha_i\bar{\alpha}_i} \ldots
F_{\alpha_p\bar{\alpha_p}}{\rm (tail)} \, .
\end{equation}
We can manipulate its coefficient, because we can add a ``derived''
DUY condition:
\begin{equation}
o_i \, F_{\alpha_1\bar{\alpha}_1} \ldots (F_{\alpha_i\bar{\alpha_i}} +
{\cal D}_{(1)} + {\cal D}_{(2)} +
\cdots) \ldots F_{\alpha_p\bar{\alpha}_p}{\rm (tail)} = 0 \, ,
\end{equation}
for each 1-loop, which will contribute a factor $o_i$ to the term under
consideration.  This
 $o_i$ can be considered
as an extra degree of freedom, which can be adjusted to make the coefficient of
the
1-loop term in the equations of motion zero. Obviously this adjustment has
implications at higher orders.  If we started
from a term with $p$ 1-loops, each term in the deformed DUY condition at higher
order contains $p-1+r$ 1-loops, where
$r$ is the number of 1-loops in the DUY term.  So if $p \neq 1$ or $r \neq 0$,
we still have a term
with 1-loops and can absorb the extra factor $o_i$ by again adding a new
``derived'' DUY condition.

\begin{table}
\centering
\begin{tabular}{|c|c|}
\hline
Order     & Maximum number \\
          & of 1-loops \\
\hline\hline
$\alpha'$   & 1 \\
$\alpha'{}^2$ & 1 \\
$\alpha'{}^3$ & 0 \\
$\alpha'{}^4$ & 0 \\
\hline
\end{tabular}
\caption{Terms with more 1-loops than indicated are not considered.}
\label{oneloops}
\end{table}

Therefore we only have to be careful if we eventually end up with a term
without 1-loops.
As we have seen in section \ref{order1}, $d^3_{0,0,0}=0$, so at order $\alpha'$
there are no DUY terms
without 1-loops ($r \neq 0$).
So if we want to lower the number of 1-loops by substituting a DUY condition,
we end up at least 2 orders
of $\alpha'$ higher.  Since the highest order we hope to study is $\alpha'{}^4$,
at that order we
do not have to consider 1-loop terms. Also at order $\alpha'{}^3$, we do not have
to consider 1-loop terms,
because they will only influence terms without 1-loops for the first time at
order $\alpha'{}^5$ --- two orders
higher.  Continuing this reasoning, we obtain table \ref{oneloops}.

Concluding, we never consider terms with more than one 1-loop, so there
is only one extra unknown per 1-loop term, which we will call
$o^n_{s_1,s_2,s_3}$ after
that 1-loop term.

As a next step, we want to get rid of the $o$-coefficients altogether.
The reader can convince himself that, provided one
uses the priority rule described in the next paragraph, there is no loss of
generality if one only introduces
the $o$-factors after the elimination process, instead of
before\footnote{Basically, this is permissible because the identities
that one uses to eliminate the dependent 1-loop terms, still apply when the
1-loop is replaced by a higher order DUY deformation
piece. So at that higher order the o-coefficients will hook up together in the
same groups as in the coefficients of the
independent 1-loop terms at lower order}. But now things get very easy because
every independent 1-loop term will have
one and only one $o$-factor, that will be adjusted to get the coefficient zero.
In appendix \ref{order2} the $o$-coefficients
are implicit.
\end{enumerate}

\item Calculate the independent terms in the equations of motion.  When
eliminating a term from
an equation, its coefficient has to be appropriately transferred to the other
terms in the equation.
The priority rule reads: always try to keep terms with as many 1-loops as
possible.  This means that a term will never
transfer its coefficient to a term with less 1-loops.  So, if we do not have to
consider the coefficients of terms
with a certain number of 1-loops at the end, we might as well not construct
them in the first place and leave
them out throughout the whole calculation.

\item Because we are searching for solutions, the equations of motion have to be
zero.  Since we now have
independent terms, their coefficients separately have to be zero. From this set
of equations, we can
solve for the unknowns $l^n_{s_1,s_2,s_3}$ and $d^n_{s_1,s_2,s_3}$.
\end{enumerate}

\section{Uncharted territory: the $\alpha'{}^3$ corrections}

Pushing our method to order $ \alpha'{}^3$, we enter largely uncharted
territory. At the same time, our approach enters a new level of complexity: the
most general deformation of the lagrangian at this order consists of no less
than 36 terms while the most general deformation of the stability condition at
this order, consistent with our assumptions, counts 27 terms. Using the same
strategy as at lower orders, but now largely relying on our program, we arrive
at the following result (see appendix \ref{lagduy3} for the terms themselves):
\begin{gather}
-l^5_{0,0,1} =  - l^5_{0,0,3} = 2 \; l^5_{1,0,4} = 2 \; l^5_{1,0,6} =
-8 l^5_{1,1,4} = l^5_{1,4,30} = - l^5_{1,4,58} = 2 \; l^5_{0,1,1} = \lambda  \nonumber \\
l^5_{0,0,0} = l^5_{0,0,2} = l^5_{0,1,0} = l^5_{1,1,3} = l^5_{1,4,47} = 0 \, ,
\label{lagsolution}
\end{gather}
with
\begin{equation}
\label{lagfactor}
\lambda  = d^5_{0,0,4} +d^5_{1,0,12} \, .
\end{equation}
As for the DUY deformation:
\begin{gather}
d^5_{0,0,2} = - d^5_{0,0,4} \nonumber \\
d^5_{0,0,1} = d^5_{0,0,3} = d^5_{0,1,x} = d^5_{1,1,x} = d^5_{2,0,x} = 0 \nonumber \\
- 2 \; d^5_{0,0,0} = 2 \; d^5_{0,0,5} = d^5_{0,0,4} + 7 \; d^5_{1,0,12} \nonumber \\
2 \; d^5_{1,0,13} = 2 d^5_{1,0,16} = 2 d^5_{1,0,19} = 2 d^5_{1,0,22} = d^5_{0,0,4} +
3 d^5_{1,0,12} \nonumber \\
d^5_{1,0,14} = d^5_{1,0,15} = d^5_{1,0,17} = d^5_{1,0,18} = d^5_{1,0,20} = d^5_{1,0,21}
= d^5_{1,0,23} = d^5_{1,0,12} \, .
\end{gather}
It turns out that 23 terms in the deformation of the
lagrangian can be removed through field redefinitions, so we did not list
them in eq.\ \eqref{lagsolution}. The terms which are left
are insensitive to field redefinitions.  Note that the solution for the lagrangian
is unique up to a multiplicative factor; the $\lambda $ in eq.\ \eqref{lagfactor}
can be freely chosen by juggling with $d^5_{0,0,4}$ and/or $d^5_{1,0,12}$ in the DUY condition.

Extending our assumption about the 1-loops in the DUY condition,
all terms with ``subloops'' turn out to be vanishing.

\section{Discussion and conclusions}\label{complex}

In this paper we started from solutions to Yang-Mills which define a
stable holomorphic bundle, eqs.\ (\ref{hol}) and (\ref{duy}). In the
context of D-brane physics this corresponds to BPS configurations in the
limit where the magnetic background fields are small. Subsequently we
investigated the deformations of
both the action and the stability condition.
As deformations we allowed for arbitrary, independent powers of the fieldstrength and
covariant derivatives thereof.
Requiring that these configurations
solve the (deformed) equations of motion largely fixes the allowed deformations.

Novel compared to the analysis in \cite{nieuw}, is the necessity to include
terms in the action with derivatives acting on the fieldstrengths. Indeed
an initial attempt which considered only deformations polynomial
in the fieldstrength failed at order $\alpha'{}^4$!

In the present paper we
studied the deformation through order $\alpha'{}^3$. Through order
$\alpha'{}^2$ we reproduce the well-known results, eqs.\ (\ref{lag0}) and
(\ref{lag2def}). The stability condition gets deformed as in eq.\
(\ref{duy2def}).

In the previous section we pushed our method to the next order. At
order $ \alpha'{}^3$ we found a one parameter family of allowed deformations,
\begin{eqnarray}
{\cal L}_{(3)}&=&-\lambda \,\tr\Big(
F_{\mu_1\mu_2}F_{\mu_2\mu_3}F_{\mu_3\mu_4}F_{\mu_5\mu_1}F_{\mu_4\mu_5}+
F_{\mu_1\mu_2}F_{\mu_4\mu_5}F_{\mu_2\mu_3}F_{\mu_5\mu_1}F_{\mu_3\mu_4}-
\nonumber\\
&&\frac 1 2
F_{\mu_1\mu_2}F_{\mu_2\mu_3}F_{\mu_4\mu_5}F_{\mu_3\mu_1}F_{\mu_5\mu_4}-
\nonumber\\
&&\frac 1 2
\left(D_{\mu_1}F_{\mu_2\mu_3}\right)\left(D_{\mu_1}F_{\mu_3\mu_4}\right)
F_{\mu_5\mu_2}F_{\mu_4\mu_5} -\frac 1 2
\left(D_{\mu_1}F_{\mu_2\mu_3}\right)F_{\mu_5\mu_2}\left(D_{\mu_1}
F_{\mu_3\mu_4}\right)F_{\mu_4\mu_5} +\nonumber \\
&&\frac 1 8
\left(D_{\mu_1}F_{\mu_2\mu_3}\right)F_{\mu_4\mu_5}\left(D_{\mu_1}
F_{\mu_3\mu_2}\right)F_{\mu_5\mu_4} -\nonumber \\
&&\left(D_{\mu_5}F_{\mu_1\mu_2}\right)F_{\mu_3\mu_4}\left(D_{\mu_1}
F_{\mu_2\mu_3}\right)F_{\mu_4\mu_5}+
F_{\mu_1\mu_2}\left(D_{\mu_1}F_{\mu_3\mu_4}\right)\left(D_{\mu_5}
F_{\mu_2\mu_3}\right)F_{\mu_4\mu_5}\Big),
\label{l3finaal}
\end{eqnarray}
with $\lambda \in\IR$. We omitted terms in eq.\ (\ref{l3finaal})
which can be removed through field redefinitions. The terms
appearing in eq.\ (\ref{l3finaal}) are inert under field redefinitions.

Let us now compare our result to the existing literature. Perhaps the cleanest
calculation can be found in \cite{bilal2}. There, a detailed analysis of the four-point
open superstring amplitude was performed. This was matched to the two-derivative terms at
order $\alpha'{}^3$ in the effective action. The result of \cite{bilal2} for these terms,
in our conventions, reads,
\begin{eqnarray}
{\cal L}_{(3)}^{der}&=& \frac{\zeta (3)}{(2\pi)^3}\tr\Big(
[F_{\mu\nu},D_\lambda F_{\sigma\mu}][D_\lambda F_{\nu\rho},F_{\rho\sigma}]+
[F_{\mu\nu},D_\lambda F_{\sigma\rho}][D_\lambda F_{\nu\rho},F_{\mu\sigma}] \nonumber\\
&&-\frac 1 2
[F_{\mu\nu},D_\lambda F_{\rho\sigma}][D_\lambda F_{\mu\nu},F_{\rho\sigma}]
\Big).\label{bilres}
\end{eqnarray}
Passing to the basis for the independent terms we chose at this order, one finds
that eq.\ (\ref{bilres}), modulo FR changeable terms,
exactly reproduces the derivative terms in eq.\ (\ref{l3finaal}) with,
\begin{eqnarray}
\lambda =- \frac{2\zeta(3)}{\pi^3},\label{valueS}
\end{eqnarray}
thereby fixing our free parameter.
In addition, the change of basis yields terms without derivatives as well. Indeed,
we get $l^5_{0,0,0}=0$, $l^5_{0,0,1}=-3\lambda /4$, $l^5_{0,0,2}=-\lambda /2$,
$l^5_{0,0,3}=-3\lambda /4$, $l^5_{0,1,0}=\lambda /8$ and  $l^5_{0,1,1}=3\lambda /8$. This does not agree with our
result in eq.\ (\ref{l3finaal}). This is not surprising as, in order to fully determine these terms,
the calculation of \cite{bilal2} has to be supplemented with the calculation and analysis of a
five-point open superstring amplitude.

Subsequently we turn to the calculation of the 
one-loop effective action of $N=4$ super Yang-Mills in
4 dimensions through operators of dimension 10 in \cite{milaan}.
As a full result at $ {\cal O}(\alpha'{}^3)$ is claimed in \cite{milaan}, we present a
detailed comparison. A particular property of $d=4$
is that only 4 of the 6 terms without derivatives are independent. This implies that the
following transformation is always possible in $d=4$,
\begin{eqnarray}
l^5_{0,0,0}&\rightarrow&l^5_{0,0,0}+\frac 3 5 l^5_{0,1,0}-\frac 1 5 l^5_{0,1,1},\nonumber\\
l^5_{0,0,1}&\rightarrow&l^5_{0,0,1}-l^5_{0,1,0}+l^5_{0,1,1} ,\nonumber\\
l^5_{0,0,2}&\rightarrow&l^5_{0,0,2}+l^5_{0,1,0}+l^5_{0,1,1} ,\nonumber\\
l^5_{0,0,3}&\rightarrow&l^5_{0,0,3}+\frac 1 5 l^5_{0,1,0}+\frac 3 5 l^5_{0,1,1} ,\nonumber\\
l^5_{0,1,0}&\rightarrow&0 ,\nonumber\\
l^5_{0,1,1}&\rightarrow&0.\label{d4rels}
\end{eqnarray}
Restricting eq.\ (\ref{l3finaal}) to four dimensions and implementing eq.\ (\ref{d4rels}) into it,
we get that the terms without derivatives are changed to
\begin{eqnarray}
l^5_{0,0,0}=- \frac{\lambda }{10},\ l^5_{0,0,1}=-\frac \lambda  2,\ l^5_{0,0,2}=\frac \lambda  2,\
l^5_{0,0,3}=- \frac{7\lambda }{10},
\ l^5_{0,1,0}=0,\ l^5_{0,1,1}=0.
\label{ourd4}
\end{eqnarray}
We now turn to eq.\ (6.8) in \cite{milaan}. It contains four terms without derivatives whose
coefficients we call, in an obvious notation,
$\hat{l}^5_{(0,0,s)}$, $s\in\{0,1,2,3\}$. The terms with derivatives
read in our conventions,
\begin{eqnarray}
{\cal L}_{(3)}^{der}&=&\tr\Big(
l^5_{1,0,3}D_{\mu_1}F_{\mu_2\mu_3}D_{\mu_1}F_{\mu_3\mu_4}F_{\mu_4\mu_5}F_{\mu_5\mu_2}+
l^5_{1,0,8}D_{\mu_1}F_{\mu_2\mu_3}F_{\mu_3\mu_4}F_{\mu_5\mu_2}D_{\mu_1}F_{\mu_4\mu_5}+ \nonumber\\
&&l^5_{1,0,4}D_{\mu_1}F_{\mu_2\mu_3}D_{\mu_1}F_{\mu_3\mu_4}F_{\mu_5\mu_2}F_{\mu_4\mu_5}+
l^5_{1,1,3}D_{\mu_1}F_{\mu_2\mu_3}D_{\mu_1}F_{\mu_3\mu_2}F_{\mu_4\mu_5}F_{\mu_5\mu_4}+ \nonumber\\
&&l^5_{1,1,6}D_{\mu_1}F_{\mu_2\mu_3}F_{\mu_3\mu_2}F_{\mu_4\mu_5}D_{\mu_1}F_{\mu_5\mu_4}+
l^5_{1,1,7}D_{\mu_1}F_{\mu_2\mu_3}D_{\mu_1}F_{\mu_4\mu_5}F_{\mu_3\mu_2}F_{\mu_5\mu_4}\Big),
\label{milac}\end{eqnarray}
where, modulo an overall multiplicative constant, \cite{milaan} gives,
\begin{eqnarray}
l^5_{1,0,3}=l^5_{1,0,4}=l^5_{1,0,8}=-\frac \lambda  4 ,\
l^5_{1,1,3}= l^5_{1,1,6}=l^5_{1,1,7}= \frac{\lambda }{16}.
\end{eqnarray}
Before comparing, we need to rewrite the $(1,0,3)$, $(1,0,8)$, $(1,1,6)$ and $(1,1,7)$
terms in our basis. Ignoring the FR removable terms, we get the following conversion table,
\begin{eqnarray}
l^5_{0,0,0}&\rightarrow&l^5_{0,0,0}+l^5_{1,0,3} ,\quad
l^5_{0,0,1}\rightarrow l^5_{0,0,1}+l^5_{1,0,3}+2l^5_{1,0,8},\nonumber\\
l^5_{0,0,2}&\rightarrow&l^5_{0,0,2}+l^5_{1,0,3}+2l^5_{1,0,8} ,\quad
l^5_{0,0,3}\rightarrow l^5_{0,0,3}+3l^5_{1,0,3},\nonumber\\
l^5_{0,1,0}&\rightarrow&l^5_{0,1,0}-\frac 1 2 l^5_{1,0,3}+4l^5_{1,1,6} ,\quad
l^5_{0,1,1}\rightarrow l^5_{0,1,1}-l^5_{1,0,3}+2l^5_{1,1,7},\nonumber\\
l^5_{1,0,4}&\rightarrow&l^5_{1,0,4}-2l^5_{1,0,3}-l^5_{1,0,8} ,\quad
l^5_{1,0,6}\rightarrow l^5_{1,0,6}-l^5_{1,0,3}-l^5_{1,0,8},\nonumber\\
l^5_{1,1,3}&\rightarrow&l^5_{1,1,3}-l^5_{1,1,6} ,\quad
l^5_{1,1,4}\rightarrow l^5_{1,1,4}+\frac 3 8 l^5_{1,0,3}-\frac 1 2 l^5_{1,1,7},\nonumber\\
l^5_{1,4,30}&\rightarrow&l^5_{1,4,30} -4l^5_{1,0,3},\quad
l^5_{1,4,47}\rightarrow l^5_{1,4,47}-l^5_{1,0,3}-4l^5_{1,1,6},\nonumber\\
l^5_{1,4,58}&\rightarrow&l^5_{1,4,58} +4l^5_{1,0,3}.
\end{eqnarray}
Upon implementing this in eq.\ (\ref{milac}) and the subsequent elimination of
the $(0,1,0)$ and $(0,1,1)$ terms using eq.\ (\ref{d4rels}),
we obtain an action of the form eq.\ (\ref{l3finaal}) where the derivative terms coincide exactly!
For the terms without derivatives, we get
\begin{eqnarray}
&&l^5_{0,0,0}=\hat{l}^5_{0,0,0}- \frac{\lambda }{10},\quad
l^5_{0,0,1}=\hat{l}^5_{0,0,1}- \frac{3\lambda }{4},\quad
l^5_{0,0,2}=\hat{l}^5_{0,0,2}, \nonumber\\
&&l^5_{0,0,3}=\hat{l}^5_{0,0,3}- \frac{9\lambda }{20},\quad
l^5_{0,1,0}=l^5_{0,1,1}=0.
\end{eqnarray}
Matching this to our result eq.\ (\ref{ourd4}), we find that we need,
\begin{eqnarray}
\hat{l}^5_{0,0,0}=0,\quad
\hat{l}^5_{0,0,1}= \frac{\lambda }{4},\quad
\hat{l}^5_{0,0,2}=\frac \lambda  2,\quad
\hat{l}^5_{0,0,3}=- \frac{5\lambda }{20}.
\end{eqnarray}
One easily checks that this does {\em not} agree with \cite{milaan}!

Finally, there is the direct calculation in \cite{kit}. We verified that the
terms with derivatives do again coincide with our derivative terms. However the
precise structure of the terms without derivatives in \cite{kit} remains obscure.
We compared various readings of \cite{kit} and always found disagreement with our
result as well as with the result in \cite{milaan}.

A consequence of all this is that we can be very confident about the derivative
terms in eq.\ (\ref{l3finaal}). It agrees perfectly with direct calculations, \cite{bilal2},
\cite{kit} and the $d=4$ $N=4$ super Yang-Mills effective action calculation \cite{milaan}.
In addition, the precise comparison of eq.\ (\ref{l3finaal}) to \cite{bilal2} fixes the free
parameter $\lambda $ as in eq.\ (\ref{valueS}). Concerning the terms without derivatives, no agreement
exists between our calculation and the result in \cite{milaan}.
This shows that in general one should not expect a direct relation between
the tree-level open string effective action, which is calculated here, 
and the quantum super Yang-Mills effective action (for a 
more detailed discussion, we refer to \cite{buch}), which is studied in \cite{milaan}. 
In fact, already in the abelian case, it is 
known that the $F^8$ term in the one-loop $N=4$ super Yang-Mills effective action is different
in structure, \cite{buch}, from the $F^8$ term in the Born-Infeld action \cite{ft}.
We are confident that eq.\ (\ref{l3finaal}),
modulo field redefinitions, together with
eq.\ (\ref{valueS}) is the non-abelian Born-Infeld action at $ {\cal O}(\alpha'{}^3)$. Indeed, the only
ansatz we made is that configurations satisfying eq.\ (\ref{hol}) and some deformation of
eq.\ (\ref{duy}) solve the equations of motion. In both the abelian limit and in the limit that
$2\pi\alpha'F$ is small, such solutions are known to represent BPS configurations of D-branes.
It is hard to conceive that such configurations would cease to exist away from these limits.
In other words, if these BPS configurations exist in the non-abelian case where
$2\pi\alpha'F$ is not necessarily small, then $ {\cal L}=
{\cal L}_{(0)}+{\cal L}_{(2)}+{\cal L}_{(3)}$,
with ${\cal L}_{(0)}$, ${\cal L}_{(2)}$ and ${\cal L}_{(3)}$ given in eqs.\ (\ref{lag0}),
(\ref{lag2def}) and (\ref{l3finaal}) resp., should be the non-abelian Born-Infeld action through
order $\alpha'{^3}$!

A very interesting and strong check of our result would follow the program set up in \cite{HT} and
further developed in \cite{STT}. There the mass spectrum of D-branes at angles is calculated.
Upon T-dualizing, this corresponds to stacks of D-branes in the presence of constant magnetic
background fields. The non-abelian Born-Infeld action should reproduce this spectrum. In particular,
the string theoretic calculation shows that the spectrum of the off-diagonal gauge fields receives
only contributions at even powers in $\alpha'$. This means that the contribution to the spectrum
of the terms without derivatives in eq.\ (\ref{l3finaal}) should exactly cancel against the contributions
coming from the terms with derivatives! Work in this direction is in progress\footnote{Note added in proof:
In meanwhile this program has been carried out in \cite{KS}. The result in eq. (\ref{l3finaal}) indeed correctly 
reproduces the spectrum, while the results of \cite{kit} and \cite{milaan} do not.}.

The result in eq.\ (\ref{l3finaal}) is not sufficient to make all order predictions about the structure
of the non-abelian Born-Infeld action. Presently, our software has been optimized to tackle the Born-Infeld
action at the next order. Hopefully this will shed some light on the all-order structure of the non-abelian
Born-Infeld action.

The fact that we obtain a one-parameter family of allowed deformations at order $\alpha'{}^3$ suggests
that supersymmetry alone is not sufficient to fix the full non-abelian Born-Infeld action. In this context
it would be most interesting to push the analysis of \cite{goteborg} one order higher.
Note that it is very fortunate that we did obtain a one parameter family of solutions. Indeed,
it is clear that, in units where $2\pi\alpha'=1$, our method can only fix coefficients in terms of rational
numbers. From open superstring amplitudes, one expects that the coefficients at order $\alpha'{}^n$ are given by
$\zeta(n)/\pi^n$ times a rational number. For $n$ even this is a rational number while for $n$ odd it is not!
So it would not be too surprising that our method would fix the deformation completely at even orders in
$\alpha'$, while giving a one-parameter family of deformations at odd orders.

\bigskip

\acknowledgments

\bigskip

We thank Arkady Tseytlin, Eric Bergshoeff, Adel Bilal, Pieter de Groen, Mees de Roo, Yoshihisa Kitazawa, Kasper
Peeters, Andrea Refolli, Walter Troost and Mariah Zamaklar
for useful discussions. In particular, we thank Lies De Foss\'e and Jan Troost
for collaboration in the initial stages of this work and for numerous stimulating suggestions.
P.K. and A.S. are
supported in part by the
FWO-Vlaanderen and in part by the European Commission RTN programme
HPRN-CT-2000-00131, in which the authors are associated to the university
of Leuven.

\vspace{5mm}

\appendix

\section{Notations and conventions}
\label{Conventions}
Our metric is Euclidean. Indices denoted
by $\mu $, $\nu$, ... run from 1
to $2p$ and those denoted by
$\alpha $, $\beta$, ... run from 1 to $p$. We choose
anti-hermitian matrices for the $u(n)$ generators. The fieldstrength and
covariant
derivative are given by
\begin{eqnarray}
F_{\mu\nu}&=&\partial_\mu A_\nu-\partial_\nu A_\mu+{[}A_\mu,A_\nu{]},
\nonumber\\
D_\mu\cdot&=&\partial_\mu\cdot+{[}A_\mu,\cdot{]}.
\end{eqnarray}

Instead of using real spatial coordinates $x^{\mu}$, ${\mu}\in\{1,\cdots,2p\}$, we
will often use complex coordinates $z^\alpha $, $\alpha \in\{1,\cdots
p\}$,
\begin{eqnarray}
z^\alpha \equiv \frac{1}{\sqrt 2}\left(x^{2\alpha -1}+ix^{2\alpha
}\right),\quad
\bar z^{\bar\alpha} \equiv \frac{1}{\sqrt 2}
\left(x^{2\alpha -1}-ix^{2\alpha }\right).
\label{cc}
\end{eqnarray}
As we work in flat space, the metric is $g_{\alpha \beta}=g_{\bar\alpha
\bar\beta}=0$, $g_{\alpha \bar\beta}=\delta_{\alpha \bar\beta}$.

Finally, we explain our strategy in classifying terms in the action, the
equations of  motion, ... Each term contains a number of derivatives $n(D)$ and
fieldstrengths $n(F)$.  Terms are classified according to (hierarchically):
\begin{enumerate}
\item {\em order} $n$:  for terms in the lagrangian the order can
be calculated by:
\begin{equation}
\label{order}
n = n(F) + \frac{n(D)}{2} \, .
\end{equation}
This corresponds in fact to order $\alpha'{}^{(n-2)}$.
For terms in the equations of motion, we define the order as the order of the
terms in the lagrangian from which the terms are
derived.  We can still use formula \eqref{order} if we count the free index as
an extra derivative.
\item {\em superstructure} $s_1$: within a certain order, one can classify the
terms according to the number of derivatives.
In the abelian case, the different superstructures do not communicate; in the
non-abelian case there are identities connecting them. They read in their most
general form:
\begin{equation}
D_1 \ldots D_k [D_{k+1},D_{k+2}] D_{k+3} \ldots D_n F_{l_1 l_2} = D_1 \ldots
D_k [F_{k+1;k+2}, D_{k+3} \ldots D_n F_{l_1 l_2}] \,
.
\end{equation}
In the rest of the article we will use the shorthand notation
$[D,D]\cdot=[F,\cdot]$.
\item {\em index structure} $s_2$:
it is convenient to use an example to explain this. At $ \alpha'{}^2$, we have
the following four terms without derivatives:
\begin{subequations}
\begin{eqnarray}
\label{F4ex1}
&l^4_{0,0,0} \; F_{\mu_1\mu_2}F_{\mu_2\mu_3}F_{\mu_3\mu_4}F_{\mu_4\mu_1} \, ,
&l^4_{0,0,1} \; F_{\mu_1\mu_2}F_{\mu_2\mu_3}F_{\mu_4\mu_1}F_{\mu_3\mu_4} \, ,
\\
\label{F4ex2}
&l^4_{0,1,0} \; F_{\mu_1\mu_2}F_{\mu_2\mu_1}F_{\mu_3\mu_4}F_{\mu_4\mu_3} \, ,
&l^4_{0,1,1} \; F_{\mu_1\mu_2}F_{\mu_3\mu_4}F_{\mu_2\mu_1}F_{\mu_4\mu_3}.
\end{eqnarray}
We note that the first two contain a loop\footnote{An $n$-loop
contains $n$ fieldstrengths, traced over the Lorentz indices and
disregarding any ordering. } with 4 $F$'s, while the latter two
contain two loops with 2 $F$'s each. For the first two we take
$s_2=0$, while the latter we label by $s_2=1$. So $s_2$ will
distinguish terms with different loop structure. When classifying
terms with derivatives e.g. in the equations of motion, one will
also have ``chains''. Those are Lorentz contracted sets of
fieldstrengths with two indices contracted with derivatives or one
index contracted with a derivative and the other one a free index.
We will call the latter type ``a free index chain''.  Two examples
containing chains, one from the lagrangian and one from the
equations of motion are:
\begin{eqnarray}
\label{F4ex3}
&& l^4_{1,2,1} \;
\left(D_{\mu_4}D_{\mu_1}F_{\mu_1\mu_2}\right)F_{\mu_3\mu_4}F_{\mu_2\mu_3} \\
\label{F4ex4}
&& e^4_{0,0,0} \; \left(D_{\bar{\beta}}F_{\alpha_1\bar{\alpha}_2}\right)F_{\alpha_2\bar{\alpha}_3
}F_{\alpha_3\bar{\alpha}_1}
\end{eqnarray}
\end{subequations}
Note that the index structure of a term can be elegantly represented by a
graph: see figure \ref{indexgraph}.
\FIGURE{
\centering
\setlength{\unitlength}{1.5mm}
\begin{picture}(80,45)(0,0)
\put(15,40){\line(1,0){10}}
\put(15,30){\line(1,0){10}}
\put(15,30){\line(0,1){10}}
\put(25,30){\line(0,1){10}}
\multiput(15,40)(10,0){2}{\usebox{\Fcirc}}
\multiput(15,30)(10,0){2}{\usebox{\Fcirc}}
\put(20,25){\makebox(0,0){Graph \eqref{F4ex1}}}
\multiput(55,40)(10,0){2}{\usebox{\Fcirc}}
\multiput(55,30)(10,0){2}{\usebox{\Fcirc}}
\qbezier(55,40)(60,43)(65,40)
\qbezier(55,40)(60,37)(65,40)
\qbezier(55,30)(60,33)(65,30)
\qbezier(55,30)(60,27)(65,30)
\put(60,25){\makebox(0,0){Graph \eqref{F4ex2}}}
\put(0,10){\line(1,0){40}}
\multiput(0,10)(10,0){5}{\usebox{\Fcirc}}
\put(0,8){\makebox(0,0)[t]{$\scriptstyle D$}}
\put(40,8){\makebox(0,0)[t]{$\scriptstyle D$}}
\put(20,5){\makebox(0,0){Graph \eqref{F4ex3}}}
\put(55,15){\line(1,0){10}}
\put(55,13){\makebox(0,0)[t]{$\scriptstyle D$}}
\put(65,13){\makebox(0,0)[t]{$\times$}}
\put(55,5){\line(1,0){10}}
\multiput(55,15)(10,0){2}{\usebox{\Fcirc}}
\multiput(55,5)(10,0){2}{\usebox{\Fcirc}}
\put(60,12.5){\usebox{\Fcirc}}
\put(55,5){\line(2,3){5}}
\put(65,5){\line(-2,3){5}}
\put(60,0){\makebox(0,0){Graph \eqref{F4ex4}}}
\end{picture}
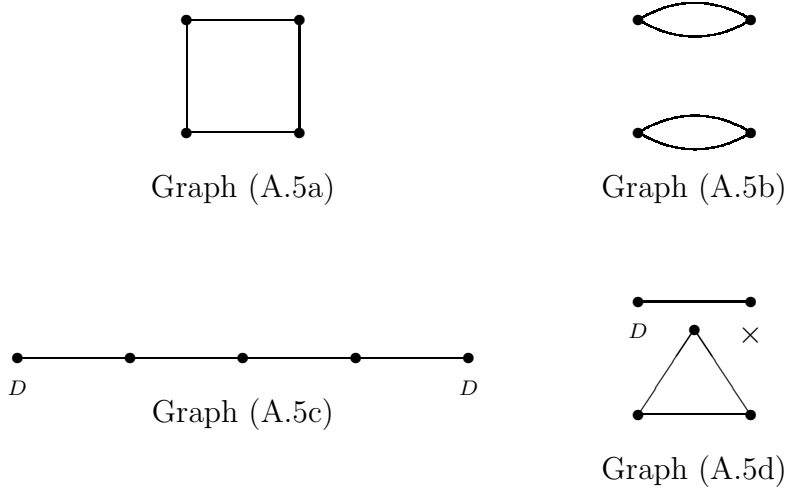
\caption{Graphs for the terms of \eqref{F4ex1}-\eqref{F4ex4}. The free index is
denoted by a $\times$. Note
that in term \eqref{F4ex4} the free index chain contains zero $F$'s.}
\label{indexgraph}
}
\item {\em term number} $s_3$.  Other classifications could be thought of, like the
derivative structure, which denote on
which terms the derivatives act.  In the non-abelian case the order of the
$F$'s is also important.  We take these two
together and just number the different terms within an index structure.
For its particular value, we take that which is used in the program. As the
program starts by writing down all possible terms and subsequently eliminates
the dependent ones through partial integration, Bianchi identities, etc. the
concrete values of $s_3$ will not necessarily be in numerical order.
\end{enumerate}

Concluding, a term in the lagrangian will be labelled by $l^n_{s_1,s_2,s_3}$. In
a similar way we label terms in the equations of motion by $e^n_{s_1,s_2,s_3}$,
those in the field redefinitions by $f^n_{s_1,s_2,s_3}$ and those in the
deformations of the DUY condition, eq.\ (\ref{duy}), by
$d^n_{s_1,s_2,s_3}$.

\section{The equations of motion at order $ \alpha'{}^2$}
\label{order2}

In this appendix, we give all contributions to the equations
of motion at order $ \alpha'{}^2$. They follow in a straightforward way from
eq.\ (\ref{lag2}). In addition to this, we take the effects of deforming
eq.\ (\ref{duy}) into account. At this order, the most general deformation of
the stability condition consistent with our assumption stated in section \ref{leading}, reads
as
\begin{equation}
 F_{\alpha\bar\alpha}+ {\cal D}_{(2)} + {\cal O}(\alpha'{}^3)= 0\, ,
\end{equation}
with
\begin{eqnarray}
{\cal D}_{(2)} & = & d^4_{0,0,0}F_{\alpha_1\bar{\alpha}_2}F_{\alpha_2\bar{\alpha}_3}
F_{\alpha_3\bar{\alpha}_1} +
d^4_{0,0,1}F_{\alpha_1\bar{\alpha}_2}F_{\alpha_3\bar{\alpha}_1}
F_{\alpha_2\bar{\alpha}_3}+
\nonumber\\
&&d^4_{1,0,4}\left(D_{\alpha_1}F_{\alpha_2\bar{\alpha}_3}\right)
\left(D_{\bar{\alpha}_1}
F_{\alpha_3\bar{\alpha}_2}\right) +
d^4_{1,0,5}\left(D_{\bar{\alpha}_1}F_{\alpha_2\bar{\alpha}_3}\right)
\left(D_{\alpha_1}F_{\alpha_3\bar{\alpha}_2}\right).\label{duy2}
\end{eqnarray}
The leading term in the equations of motion,
$D_{\bar\beta}F_{\alpha\bar\alpha}=0$, vanished because of eq.\ (\ref{duy}).
If we now deform eq.\ (\ref{duy}) as in eq.\ (\ref{duy2}), we will
induce further contributions to the equations of motion.

Below, we list the contribution to the equations of motion following from
$ {\cal L}_{(2)}$, eq.\ (\ref{lag2}) and the corrections following from ${\cal D}_{(2)}$,
eq.\ (\ref{duy2}). We list them below in hierarchical order
(as was explained in appendix \ref{Conventions}) and omit terms with more than one 1-loop (see table \ref{oneloops}).

\subsubsection*{Superstructure 0: \#derivatives: 1, \#$F$s: 3}

Structure 0: Free index chain length: 0 Chains: 0 Loops: 3
\begin{eqnarray}
& e^4_{0,0,0}:
&\left(D_{\bar{\beta}}F_{\alpha_1\bar{\alpha}_2}\right)F_{\alpha_2
\bar{\alpha}_3}F_{\alpha_3\bar{\alpha}_1} \nonumber \\
& &+2 \; l^4_{0,0,1}- \; d^4_{0,0,0}-2 \; l^4_{1,2,1} \nonumber \\
& e^4_{0,0,1}:
&\left(D_{\bar{\beta}}F_{\alpha_1\bar{\alpha}_2}\right)F_{\alpha_3
\bar{\alpha}_1}F_{\alpha_2\bar{\alpha}_3} \nonumber \\
& &+4 \; l^4_{0,0,0}- \; d^4_{0,0,1}+2 \; l^4_{1,2,1} \nonumber \\
& e^4_{0,0,2}:
&F_{\alpha_1\bar{\alpha}_2}\left(D_{\bar{\beta}}F_{\alpha_3\bar{\alpha}_1}
\right)F_{\alpha_2\bar{\alpha}_3} \nonumber \\
& &-2 \; l^4_{0,0,1}- \; d^4_{0,0,1}-16 \; l^4_{0,1,1}-8 \; l^4_{0,1,0}
\nonumber \\
& e^4_{0,0,3}:
&F_{\alpha_1\bar{\alpha}_2}\left(D_{\bar{\beta}}F_{\alpha_2\bar{\alpha}_3}
\right)F_{\alpha_3\bar{\alpha}_1} \nonumber \\
& &+6 \; l^4_{0,0,1}- \; d^4_{0,0,0}+16 \; l^4_{0,1,1}+8 \; l^4_{0,1,0}
\nonumber \\
& e^4_{0,0,4}:
&F_{\alpha_1\bar{\alpha}_2}F_{\alpha_2\bar{\alpha}_3}\left(D_{\bar{\beta}}
F_{\alpha_3\bar{\alpha}_1}\right) \nonumber \\
& &+4 \; l^4_{0,0,0}- \; d^4_{0,0,0}+2 \; l^4_{1,2,1}+4 \; l^4_{0,0,1}+16 \;
l^4_{0,1,1}+8 \; l^4_{0,1,0} \nonumber \\
& e^4_{0,0,5}:
&F_{\alpha_1\bar{\alpha}_2}F_{\alpha_3\bar{\alpha}_1}\left(D_{\bar{\beta}}
F_{\alpha_2\bar{\alpha}_3}\right) \nonumber \\
& &-2 \; l^4_{0,0,1}- \; d^4_{0,0,1}-2 \; l^4_{1,2,1}-16 \; l^4_{0,1,1}-8 \;
l^4_{0,1,0}
\end{eqnarray}
Structure 1: Free index chain length: 0 Chains: 0 Loops: 2 1
\begin{eqnarray}
& e^4_{0,1,0}:
&\left(D_{\bar{\beta}}F_{\alpha_1\bar{\alpha}_2}\right)F_{\alpha_2
\bar{\alpha}_1}F_{\alpha_3\bar{\alpha}_3} \nonumber \\
& &+ \; l^4_{1,2,1}- \; l^4_{1,2,6}-4 \; l^4_{2,1,23}
 \nonumber \\
& e^4_{0,1,1}:
&\left(D_{\bar{\beta}}F_{\alpha_1\bar{\alpha}_2}\right)F_{\alpha_3
\bar{\alpha}_3}F_{\alpha_2\bar{\alpha}_1} \nonumber \\
& &-2 \; l^4_{0,0,1}- \; l^4_{1,2,1} \nonumber \\
& e^4_{0,1,2}:
&F_{\alpha_1\bar{\alpha}_2}\left(D_{\bar{\beta}}F_{\alpha_2\bar{\alpha}_1}
\right)F_{\alpha_3\bar{\alpha}_3} \nonumber \\
& &+ \; l^4_{1,2,1}
+ \; l^4_{1,2,6}+4 \; l^4_{2,1,23}+4 \; l^4_{0,0,1}+16 \; l^4_{0,1,1}+8 \;
l^4_{0,1,0}  \nonumber \\
& e^4_{0,1,3}:
&F_{\alpha_1\bar{\alpha}_1}\left(D_{\bar{\beta}}F_{\alpha_2\bar{\alpha}_3}
\right)F_{\alpha_3\bar{\alpha}_2} \nonumber \\
& &+2 \; l^4_{0,0,1} + \; l^4_{1,2,6} +4 \; l^4_{2,1,23}
\nonumber \\
& e^4_{0,1,4}:
&F_{\alpha_1\bar{\alpha}_2}F_{\alpha_3\bar{\alpha}_3}\left(D_{\bar{\beta}}
F_{\alpha_2\bar{\alpha}_1}\right) \nonumber \\
& &-2 \; l^4_{1,2,1} -4 \; l^4_{0,0,1}-16 \; l^4_{0,1,1}-8 \;
l^4_{0,1,0} \nonumber \\
& e^4_{0,1,5}:
&F_{\alpha_1\bar{\alpha}_1}F_{\alpha_2\bar{\alpha}_3}\left(D_{\bar{\beta}}
F_{\alpha_3\bar{\alpha}_2}\right) \nonumber \\
& &+ \; l^4_{1,2,1}- \; l^4_{1,2,6}-4 \; l^4_{2,1,23}
 \nonumber \\
& e^4_{0,1,6}:
&F_{\alpha_1\bar{\alpha}_2}F_{\alpha_2\bar{\alpha}_1}\left(D_{\bar{\beta}}
F_{\alpha_3\bar{\alpha}_3}\right) \nonumber \\
& &+2 \; l^4_{0,0,1}- \; l^4_{1,2,6}-4 \; l^4_{2,1,23}+16 \; l^4_{0,1,0}
+16 \; l^4_{0,1,1}+ \; l^4_{1,2,1}
\nonumber \\
& e^4_{0,1,7}:
&F_{\alpha_1\bar{\alpha}_2}\left(D_{\bar{\beta}}F_{\alpha_3\bar{\alpha}_3}
\right)F_{\alpha_2\bar{\alpha}_1} \nonumber \\
& &-2 \; l^4_{1,2,1}-4 \; l^4_{0,0,1}
-8 \; l^4_{0,1,0} \nonumber \\
& e^4_{0,1,8}:
&\left(D_{\bar{\beta}}F_{\alpha_1\bar{\alpha}_1}\right)F_{\alpha_2
\bar{\alpha}_3}F_{\alpha_3\bar{\alpha}_2} \nonumber \\
& &+ \; l^4_{1,2,1}+ \; l^4_{1,2,6} +4 \; l^4_{2,1,23}+8 \; l^4_{0,1,0}+2 \;
l^4_{0,0,1}
\end{eqnarray}
Structure 2: Free index chain length: 0 Chains: 0 Loops: 1 1 1 \\
Terms omitted. \\
Structure 3: Free index chain length: 1 Chains: 1 Loops: 2
\begin{eqnarray}
& e^4_{0,3,0}:
&F_{\alpha_1\bar{\beta}}\left(D_{\bar{\alpha}_1}F_{\alpha_2\bar{\alpha}_3}
\right)F_{\alpha_3\bar{\alpha}_2} \nonumber \\
& &+4 \; l^4_{0,0,0}+8 \; l^4_{0,1,0} \nonumber \\
& e^4_{0,3,2}:
&\left(D_{\bar{\alpha}_1}F_{\alpha_2\bar{\alpha}_3}\right)
F_{\alpha_1\bar{\beta}}F_{\alpha_3\bar{\alpha}_2} \nonumber \\
& &+4 \; l^4_{0,0,1}+16 \; l^4_{0,1,1}+8 \; l^4_{0,1,0} \nonumber \\
& e^4_{0,3,5}:
&F_{\alpha_2\bar{\alpha}_3}\left(D_{\bar{\alpha}_1}F_{\alpha_3\bar{\alpha}_2}
\right)F_{\alpha_1\bar{\beta}} \nonumber \\
& &+4 \; l^4_{0,0,0}+16 \; l^4_{0,1,0}+4 \; l^4_{0,0,1}+16 \; l^4_{0,1,1}
\end{eqnarray}
Structure 5: Free index chain length: 2 Chains: 2 Loops: 1
\begin{eqnarray}
& e^4_{0,5,2}:
&F_{\alpha_2\bar{\beta}}F_{\alpha_1\bar{\alpha}_2}\left(D_{\bar{\alpha}_1}
F_{\alpha_3\bar{\alpha}_3}\right) \nonumber \\
& &+4 \; l^4_{0,0,1}+4 \; l^4_{0,0,0} \nonumber \\
& e^4_{0,5,3}:
&\left(D_{\bar{\alpha}_1}F_{\alpha_3\bar{\alpha}_3}\right)F_{\alpha_1\bar{
\alpha}_2}F_{\alpha_2\bar{\beta}} \nonumber \\
& &+4 \; l^4_{0,0,1}+4 \; l^4_{0,0,0}
\end{eqnarray}

\subsubsection*{Superstructure 1: \#derivatives: 3, \#$F$s: 2}

Structure 0: Free index chain length: 0 Chains: 0 0 Loops: 2
\begin{eqnarray}
& e^4_{1,0,7}:
&F_{\alpha_2\bar{\alpha}_3}\left(D_{\alpha_1}D_{\bar{\beta}}D_{\bar{\alpha}_1}
F_{\alpha_3\bar{\alpha}_2}\right) \nonumber \\
& &+4 \; l^4_{0,0,1}+16 \; l^4_{0,1,1}+8 \; l^4_{0,1,0} \nonumber \\
& e^4_{1,0,16}:
&\left(D_{\bar{\beta}}D_{\alpha_1}F_{\alpha_2\bar{\alpha}_3}\right)
\left(D_{\bar{\alpha}_1}F_{\alpha_3\bar{\alpha}_2}\right) \nonumber \\
& &- \; d^4_{1,0,4}-2 \; l^4_{0,0,1}-8 \; l^4_{0,1,0}- \; l^4_{1,2,1}
\nonumber \\
& e^4_{1,0,17}:
&\left(D_{\bar{\alpha}_1}F_{\alpha_2\bar{\alpha}_3}\right)\left(D_{\bar{\beta}}
D_{\alpha_1}F_{\alpha_3\bar{\alpha}_2}\right) \nonumber \\
& &+ \; l^4_{1,2,1}- \; d^4_{1,0,5}+2 \; l^4_{0,0,1}+8 \; l^4_{0,1,0}
\nonumber \\
& e^4_{1,0,18}:
&\left(D_{\bar{\beta}}D_{\bar{\alpha}_1}F_{\alpha_2\bar{\alpha}_3}\right)\left(
D_{\alpha_1}F_{\alpha_3\bar{\alpha}_2}\right) \nonumber \\
& &- \; d^4_{1,0,5}+ \; l^4_{1,2,1} \nonumber \\
& e^4_{1,0,19}:
&\left(D_{\alpha_1}F_{\alpha_2\bar{\alpha}_3}\right)\left(D_{\bar{\beta}}
D_{\bar{\alpha}_1}F_{\alpha_3\bar{\alpha}_2}\right) \nonumber \\
& &- \; l^4_{1,2,1}- \; d^4_{1,0,4} \nonumber \\
& e^4_{1,0,20}:
&\left(D_{\alpha_1}D_{\bar{\beta}}F_{\alpha_2\bar{\alpha}_3}\right)\left(
D_{\bar{\alpha}_1}F_{\alpha_3\bar{\alpha}_2}\right) \nonumber \\
& &+2 \; l^4_{1,2,1}+2 \; l^4_{0,0,1}+8 \; l^4_{0,1,0} \nonumber \\
& e^4_{1,0,21}:
&\left(D_{\bar{\alpha}_1}F_{\alpha_2\bar{\alpha}_3}\right)\left(D_{\alpha_1}
D_{\bar{\beta}}F_{\alpha_3\bar{\alpha}_2}\right) \nonumber \\
& &-2 \; l^4_{1,2,1}-2 \; l^4_{0,0,1}-8 \; l^4_{0,1,0}
\end{eqnarray}
Structure 1: Free index chain length: 0 Chains: 0 0 Loops: 1 1
\\ Terms omitted. \\
Structure 3: Free index chain length: 0 Chains: 1 0 Loops: 1
\begin{eqnarray}
& e^4_{1,3,4}:
&\left(D_{\alpha_1}D_{\bar{\beta}}F_{\alpha_2\bar{\alpha}_1}\right)\left(
D_{\bar{\alpha}_2}F_{\alpha_3\bar{\alpha}_3}\right) \nonumber \\
& &-2 \; l^4_{1,2,6}-8 \; l^4_{2,1,23}-4 \; l^4_{0,0,1}
- \; l^4_{1,2,1} \nonumber \\
& e^4_{1,3,5}:
&\left(D_{\bar{\alpha}_2}F_{\alpha_3\bar{\alpha}_3}\right)\left(D_{\alpha_1}D_{
\bar{\beta}}F_{\alpha_2\bar{\alpha}_1}\right) \nonumber \\
& &+2 \; l^4_{1,2,6}+8 \; l^4_{2,1,23} +4 \; l^4_{0,0,1}
+ \; l^4_{1,2,1} \nonumber \\
& e^4_{1,3,8}:
&\left(D_{\bar{\beta}}F_{\alpha_2\bar{\alpha}_1}\right)\left(D_{\bar{\alpha}_2}
D_{\alpha_1}F_{\alpha_3\bar{\alpha}_3}\right) \nonumber \\
& &-8 \; l^4_{2,1,23} -2 \; l^4_{1,2,6}+2 \; l^4_{1,2,1}
\nonumber \\
& e^4_{1,3,9}:
&\left(D_{\bar{\alpha}_2}D_{\alpha_1}F_{\alpha_3\bar{\alpha}_3}\right)\left(D_{
\bar{\beta}}F_{\alpha_2\bar{\alpha}_1}\right) \nonumber \\
& &+8 \; l^4_{2,1,23}+2 \; l^4_{1,2,6}-2 \; l^4_{1,2,1}
\nonumber \\
& e^4_{1,3,13}:
&\left(D_{\bar{\beta}}D_{\bar{\alpha}_2}D_{\alpha_1}F_{\alpha_3\bar{\alpha}_3}
\right)F_{\alpha_2\bar{\alpha}_1} \nonumber \\
& &+ \; l^4_{1,2,1}-2 \; l^4_{0,0,1} \nonumber \\
& e^4_{1,3,14}:
&F_{\alpha_1\bar{\alpha}_2}\left(D_{\bar{\beta}}D_{\alpha_2}D_{\bar{\alpha}_1}
F_{\alpha_3\bar{\alpha}_3}\right) \nonumber \\
& &-2 \; l^4_{0,0,1}- \; l^4_{1,2,1}-16 \; l^4_{0,1,1}-8 \;
l^4_{0,1,0} \nonumber \\
& e^4_{1,3,20}:
&F_{\alpha_2\bar{\alpha}_1}\left(D_{\bar{\alpha}_2}D_{\alpha_1}D_{\bar{\beta}}
F_{\alpha_3\bar{\alpha}_3}\right) \nonumber \\
& &-2 \; l^4_{1,2,6}-8 \; l^4_{2,1,23}-2 \; l^4_{0,0,1}- \; l^4_{1,2,1}
\nonumber \\
& e^4_{1,3,23}:
&\left(D_{\alpha_2}D_{\bar{\alpha}_1}D_{\bar{\beta}}F_{\alpha_3\bar{\alpha}_3}
\right)F_{\alpha_1\bar{\alpha}_2} \nonumber \\
& &+2 \; l^4_{1,2,6}+8 \; l^4_{2,1,23}+2 \; l^4_{0,0,1}
+ \; l^4_{1,2,1}
\end{eqnarray}

\subsubsection*{Superstructure 2: \#derivatives: 5, \#$F$s: 1}

Structure 0: Free index chain length: 0 Chains: 0 0 0 Loops: 1
\begin{eqnarray}
& e^4_{2,0,0}:
&\left(D_{\bar{\beta}}D_{\bar{\alpha}_2}D_{\alpha_2}D_{\bar{\alpha}_1}
D_{\alpha_1}F_{\alpha_3\bar{\alpha}_3}\right) \nonumber \\
& &+8 \; l^4_{2,1,23}-
\; l^4_{1,2,1} \nonumber \\
& e^4_{2,0,6}:
&\left(D_{\alpha_2}D_{\bar{\beta}}D_{\bar{\alpha}_2}D_{\bar{\alpha}_1}
D_{\alpha_1}F_{\alpha_3\bar{\alpha}_3}\right) \nonumber \\
& &-2 \; l^4_{1,2,6}-8 \; l^4_{2,1,23} \nonumber \\
& e^4_{2,0,48}:
&\left(D_{\bar{\alpha}_2}D_{\alpha_2}D_{\bar{\alpha}_1}D_{\alpha_1}
D_{\bar{\beta}}F_{\alpha_3\bar{\alpha}_3}\right) \nonumber \\
& &+ \; l^4_{1,2,1}+2 \; l^4_{1,2,6}+8 \; l^4_{2,1,23}
\end{eqnarray}

\section{Lagrangian and DUY deformation at order $\alpha'{}^3$}
\label{lagduy3}

In this appendix we will quote a base for the most general lagrangian and DUY deformation
at order $\alpha'{}^3$ for further use in the text.  These were calculated by our computer program following
the method described in section \ref{algorithm}.

Below we first list the lagrangian ${\cal L}_{(3)}$, with terms sensitive to field redefinitions marked by (FR).
For the numbering logic, see appendix \ref{Conventions}.

\subsubsection*{Superstructure 0: \#derivatives: 0, \#$F$s: 5}
Structure 0: Loops: 5
\begin{eqnarray}
& l^5_{0,0,0}: &F_{\mu_1\mu_2}F_{\mu_2\mu_3}F_{\mu_3\mu_4}F_{\mu_4\mu_5}F_{\mu_5\mu_1} \nonumber \\
& l^5_{0,0,1}: &F_{\mu_1\mu_2}F_{\mu_2\mu_3}F_{\mu_3\mu_4}F_{\mu_5\mu_1}F_{\mu_4\mu_5} \nonumber \\
& l^5_{0,0,2}: &F_{\mu_1\mu_2}F_{\mu_2\mu_3}F_{\mu_5\mu_1}F_{\mu_3\mu_4}F_{\mu_4\mu_5} \nonumber \\
& l^5_{0,0,3}: &F_{\mu_1\mu_2}F_{\mu_4\mu_5}F_{\mu_2\mu_3}F_{\mu_5\mu_1}F_{\mu_3\mu_4} \nonumber \\
\end{eqnarray}
Structure 1: Loops: 3 2
\begin{eqnarray}
& l^5_{0,1,0}: &F_{\mu_1\mu_2}F_{\mu_2\mu_3}F_{\mu_3\mu_1}F_{\mu_4\mu_5}F_{\mu_5\mu_4} \nonumber \\
& l^5_{0,1,1}: &F_{\mu_1\mu_2}F_{\mu_2\mu_3}F_{\mu_4\mu_5}F_{\mu_3\mu_1}F_{\mu_5\mu_4} \nonumber \\
\end{eqnarray}

\subsubsection*{Superstructure 1: \#derivatives: 2, \#$F$s: 4}

Structure 0: Chains: 0 Loops: 4
\begin{eqnarray}
& l^5_{1,0,4}: &\left(D_{\mu_1}F_{\mu_2\mu_3}\right)\left(D_{\mu_1}F_{\mu_3\mu_4}\right)F_{\mu_5\mu_2}F_{\mu_4\mu_5} \nonumber \\
& l^5_{1,0,6}: &\left(D_{\mu_1}F_{\mu_2\mu_3}\right)F_{\mu_5\mu_2}\left(D_{\mu_1}F_{\mu_3\mu_4}\right)F_{\mu_4\mu_5} \nonumber \\
\end{eqnarray}
Structure 1: Chains: 0 Loops: 2 2
\begin{eqnarray}
& l^5_{1,1,3}: &\left(D_{\mu_1}F_{\mu_2\mu_3}\right)\left(D_{\mu_1}F_{\mu_3\mu_2}\right)F_{\mu_4\mu_5}F_{\mu_5\mu_4} \nonumber \\
& l^5_{1,1,4}: &\left(D_{\mu_1}F_{\mu_2\mu_3}\right)F_{\mu_4\mu_5}\left(D_{\mu_1}F_{\mu_3\mu_2}\right)F_{\mu_5\mu_4} \nonumber \\
\end{eqnarray}
Structure 3: Chains: 2 Loops: 2
\begin{eqnarray}
& l^5_{1,3,0}: &\left(D_{\mu_3}D_{\mu_1}F_{\mu_1\mu_2}\right)F_{\mu_2\mu_3}F_{\mu_4\mu_5}F_{\mu_5\mu_4} \, {\rm (FR)} \nonumber \\
& l^5_{1,3,1}: &\left(D_{\mu_3}D_{\mu_1}F_{\mu_1\mu_2}\right)F_{\mu_4\mu_5}F_{\mu_2\mu_3}F_{\mu_5\mu_4} \, {\rm (FR)} \nonumber \\
& l^5_{1,3,2}: &\left(D_{\mu_3}D_{\mu_1}F_{\mu_1\mu_2}\right)F_{\mu_4\mu_5}F_{\mu_5\mu_4}F_{\mu_2\mu_3} \, {\rm (FR)} \nonumber \\
& l^5_{1,3,10}: &\left(D_{\mu_1}F_{\mu_1\mu_2}\right)\left(D_{\mu_3}F_{\mu_4\mu_5}\right)F_{\mu_2\mu_3}F_{\mu_5\mu_4} \, {\rm (FR)} \nonumber \\
\end{eqnarray}
Structure 4: Chains: 4
\begin{eqnarray}
& l^5_{1,4,0}: &\left(D_{\mu_5}D_{\mu_1}F_{\mu_1\mu_2}\right)F_{\mu_2\mu_3}F_{\mu_3\mu_4}F_{\mu_4\mu_5} \, {\rm (FR)} \nonumber \\
& l^5_{1,4,1}: &\left(D_{\mu_5}D_{\mu_1}F_{\mu_1\mu_2}\right)F_{\mu_2\mu_3}F_{\mu_4\mu_5}F_{\mu_3\mu_4} \, {\rm (FR)} \nonumber \\
& l^5_{1,4,2}: &\left(D_{\mu_5}D_{\mu_1}F_{\mu_1\mu_2}\right)F_{\mu_3\mu_4}F_{\mu_2\mu_3}F_{\mu_4\mu_5} \, {\rm (FR)} \nonumber \\
& l^5_{1,4,3}: &\left(D_{\mu_5}D_{\mu_1}F_{\mu_1\mu_2}\right)F_{\mu_4\mu_5}F_{\mu_2\mu_3}F_{\mu_3\mu_4} \, {\rm (FR)} \nonumber \\
& l^5_{1,4,5}: &\left(D_{\mu_5}D_{\mu_1}F_{\mu_1\mu_2}\right)F_{\mu_4\mu_5}F_{\mu_3\mu_4}F_{\mu_2\mu_3} \, {\rm (FR)} \nonumber \\
& l^5_{1,4,12}: &\left(D_{\mu_1}F_{\mu_1\mu_2}\right)\left(D_{\mu_5}F_{\mu_2\mu_3}\right)F_{\mu_3\mu_4}F_{\mu_4\mu_5} \, {\rm (FR)} \nonumber \\
& l^5_{1,4,13}: &\left(D_{\mu_1}F_{\mu_1\mu_2}\right)\left(D_{\mu_5}F_{\mu_2\mu_3}\right)F_{\mu_4\mu_5}F_{\mu_3\mu_4} \, {\rm (FR)} \nonumber \\
& l^5_{1,4,14}: &\left(D_{\mu_1}F_{\mu_1\mu_2}\right)F_{\mu_3\mu_4}\left(D_{\mu_5}F_{\mu_2\mu_3}\right)F_{\mu_4\mu_5} \, {\rm (FR)} \nonumber \\
& l^5_{1,4,15}: &\left(D_{\mu_1}F_{\mu_1\mu_2}\right)F_{\mu_4\mu_5}\left(D_{\mu_5}F_{\mu_2\mu_3}\right)F_{\mu_3\mu_4} \, {\rm (FR)} \nonumber \\
& l^5_{1,4,17}: &\left(D_{\mu_1}F_{\mu_1\mu_2}\right)F_{\mu_4\mu_5}F_{\mu_3\mu_4}\left(D_{\mu_5}F_{\mu_2\mu_3}\right) \, {\rm (FR)} \nonumber \\
& l^5_{1,4,24}: &\left(D_{\mu_1}F_{\mu_1\mu_2}\right)F_{\mu_2\mu_3}F_{\mu_3\mu_4}\left(D_{\mu_5}F_{\mu_4\mu_5}\right) \, {\rm (FR)} \nonumber \\
& l^5_{1,4,25}: &\left(D_{\mu_1}F_{\mu_1\mu_2}\right)F_{\mu_2\mu_3}\left(D_{\mu_5}F_{\mu_4\mu_5}\right)F_{\mu_3\mu_4} \, {\rm (FR)} \nonumber \\
& l^5_{1,4,27}: &\left(D_{\mu_1}F_{\mu_1\mu_2}\right)F_{\mu_3\mu_4}\left(D_{\mu_5}F_{\mu_4\mu_5}\right)F_{\mu_2\mu_3} \, {\rm (FR)} \nonumber \\
& l^5_{1,4,30}: &\left(D_{\mu_5}F_{\mu_1\mu_2}\right)F_{\mu_3\mu_4}\left(D_{\mu_1}F_{\mu_2\mu_3}\right)F_{\mu_4\mu_5}  \nonumber \\
& l^5_{1,4,47}: &F_{\mu_1\mu_2}\left(D_{\mu_1}F_{\mu_2\mu_3}\right)F_{\mu_4\mu_5}\left(D_{\mu_5}F_{\mu_3\mu_4}\right) \nonumber \\
& l^5_{1,4,58}: &F_{\mu_1\mu_2}\left(D_{\mu_1}F_{\mu_3\mu_4}\right)\left(D_{\mu_5}F_{\mu_2\mu_3}\right)F_{\mu_4\mu_5} \nonumber \\
\end{eqnarray}

\subsubsection*{Superstructure 2: \#derivatives: 4, \#$F$s: 3}

Structure 2: Chains: 3 0
\begin{eqnarray}
& l^5_{2,2,66}: &\left(D_{\mu_5}D_{\mu_1}F_{\mu_1\mu_2}\right)\left(D_{\mu_5}D_{\mu_4}F_{\mu_2\mu_3}\right)F_{\mu_3\mu_4} \, {\rm (FR)} \nonumber \\
& l^5_{2,2,91}: &\left(D_{\mu_5}D_{\mu_5}D_{\mu_1}F_{\mu_1\mu_2}\right)\left(D_{\mu_4}F_{\mu_3\mu_4}\right)F_{\mu_2\mu_3} \, {\rm (FR)} \nonumber \\
\end{eqnarray}
Structure 3: Chains: 2 1
\begin{eqnarray}
& l^5_{2,3,39}: &\left(D_{\mu_3}D_{\mu_4}D_{\mu_1}F_{\mu_1\mu_2}\right)\left(D_{\mu_5}F_{\mu_4\mu_5}\right)F_{\mu_2\mu_3} \, {\rm (FR)} \nonumber \\
& l^5_{2,3,70}: &\left(D_{\mu_4}D_{\mu_1}F_{\mu_1\mu_2}\right)\left(D_{\mu_3}F_{\mu_2\mu_3}\right)\left(D_{\mu_5}F_{\mu_4\mu_5}\right) \, {\rm (FR)} \nonumber \\
& l^5_{2,3,97}: &\left(D_{\mu_1}F_{\mu_1\mu_2}\right)\left(D_{\mu_4}F_{\mu_2\mu_3}\right)\left(D_{\mu_3}D_{\mu_5}F_{\mu_4\mu_5}\right) \, {\rm (FR)} \nonumber \\
\end{eqnarray}

\subsubsection*{Superstructure 3: \#derivatives: 6, \#$F$s: 2}

Structure 1: Chains: 2 0 0
\begin{eqnarray}
& l^5_{3,1,225}: &\left(D_{\mu_5}D_{\mu_4}D_{\mu_4}D_{\mu_1}F_{\mu_1\mu_2}\right)\left(D_{\mu_5}D_{\mu_3}F_{\mu_2\mu_3}\right) \, {\rm (FR)} \nonumber \\
\end{eqnarray}

We did not explicitly show how the FR changeable terms transform under field redefinitions, but we checked that in fact all coordinate
changes are independent, so that we can bring the coordinates of these terms to arbitrary values by choosing an appropriate field redefinition.

As for the most general DUY deformation at this order, it reads:

\subsubsection*{Superstructure 0: \#derivatives: 0, \#$F$s: 4}

Structure 0: Loops: 4
\begin{eqnarray}
& d^5_{0,0,0}: &F_{\alpha_1\bar{\alpha}_2}F_{\alpha_2\bar{\alpha}_3}F_{\alpha_3\bar{\alpha}_4}F_{\alpha_4\bar{\alpha}_1} \nonumber \\
& d^5_{0,0,1}: &F_{\alpha_1\bar{\alpha}_2}F_{\alpha_2\bar{\alpha}_3}F_{\alpha_4\bar{\alpha}_1}F_{\alpha_3\bar{\alpha}_4} \nonumber \\
& d^5_{0,0,2}: &F_{\alpha_1\bar{\alpha}_2}F_{\alpha_3\bar{\alpha}_4}F_{\alpha_2\bar{\alpha}_3}F_{\alpha_4\bar{\alpha}_1} \nonumber \\
& d^5_{0,0,3}: &F_{\alpha_1\bar{\alpha}_2}F_{\alpha_4\bar{\alpha}_1}F_{\alpha_2\bar{\alpha}_3}F_{\alpha_3\bar{\alpha}_4} \nonumber \\
& d^5_{0,0,4}: &F_{\alpha_1\bar{\alpha}_2}F_{\alpha_3\bar{\alpha}_4}F_{\alpha_4\bar{\alpha}_1}F_{\alpha_2\bar{\alpha}_3} \nonumber \\
& d^5_{0,0,5}: &F_{\alpha_1\bar{\alpha}_2}F_{\alpha_4\bar{\alpha}_1}F_{\alpha_3\bar{\alpha}_4}F_{\alpha_2\bar{\alpha}_3} \nonumber \\
\end{eqnarray}
Structure 1: Loops: 2 2
\begin{eqnarray}
& d^5_{0,1,0}: &F_{\alpha_1\bar{\alpha}_2}F_{\alpha_2\bar{\alpha}_1}F_{\alpha_3\bar{\alpha}_4}F_{\alpha_4\bar{\alpha}_3} \nonumber \\
& d^5_{0,1,1}: &F_{\alpha_1\bar{\alpha}_2}F_{\alpha_3\bar{\alpha}_4}F_{\alpha_2\bar{\alpha}_1}F_{\alpha_4\bar{\alpha}_3} \nonumber \\
& d^5_{0,1,2}: &F_{\alpha_1\bar{\alpha}_2}F_{\alpha_3\bar{\alpha}_4}F_{\alpha_4\bar{\alpha}_3}F_{\alpha_2\bar{\alpha}_1} \nonumber \\
\end{eqnarray}

\subsubsection*{Superstructure 1: \#derivatives: 2, \#$F$s: 3}

Structure 0: Chains: 0 Loops: 3
\begin{eqnarray}
& d^5_{1,0,12}: &\left(D_{\alpha_1}F_{\alpha_2\bar{\alpha}_3}\right)\left(D_{\bar{\alpha}_1}F_{\alpha_3\bar{\alpha}_4}\right)F_{\alpha_4\bar{\alpha}_2} \nonumber \\
& d^5_{1,0,13}: &\left(D_{\alpha_1}F_{\alpha_2\bar{\alpha}_3}\right)F_{\alpha_4\bar{\alpha}_2}\left(D_{\bar{\alpha}_1}F_{\alpha_3\bar{\alpha}_4}\right) \nonumber \\
& d^5_{1,0,14}: &\left(D_{\bar{\alpha}_1}F_{\alpha_2\bar{\alpha}_3}\right)\left(D_{\alpha_1}F_{\alpha_4\bar{\alpha}_2}\right)F_{\alpha_3\bar{\alpha}_4} \nonumber \\
& d^5_{1,0,15}: &F_{\alpha_2\bar{\alpha}_3}\left(D_{\alpha_1}F_{\alpha_3\bar{\alpha}_4}\right)\left(D_{\bar{\alpha}_1}F_{\alpha_4\bar{\alpha}_2}\right) \nonumber \\
& d^5_{1,0,16}: &\left(D_{\bar{\alpha}_1}F_{\alpha_2\bar{\alpha}_3}\right)F_{\alpha_3\bar{\alpha}_4}\left(D_{\alpha_1}F_{\alpha_4\bar{\alpha}_2}\right) \nonumber \\
& d^5_{1,0,17}: &F_{\alpha_2\bar{\alpha}_3}\left(D_{\bar{\alpha}_1}F_{\alpha_4\bar{\alpha}_2}\right)\left(D_{\alpha_1}F_{\alpha_3\bar{\alpha}_4}\right) \nonumber \\
& d^5_{1,0,18}: &\left(D_{\bar{\alpha}_1}F_{\alpha_2\bar{\alpha}_3}\right)\left(D_{\alpha_1}F_{\alpha_3\bar{\alpha}_4}\right)F_{\alpha_4\bar{\alpha}_2} \nonumber \\
& d^5_{1,0,19}: &\left(D_{\bar{\alpha}_1}F_{\alpha_2\bar{\alpha}_3}\right)F_{\alpha_4\bar{\alpha}_2}\left(D_{\alpha_1}F_{\alpha_3\bar{\alpha}_4}\right) \nonumber \\
& d^5_{1,0,20}: &\left(D_{\alpha_1}F_{\alpha_2\bar{\alpha}_3}\right)\left(D_{\bar{\alpha}_1}F_{\alpha_4\bar{\alpha}_2}\right)F_{\alpha_3\bar{\alpha}_4} \nonumber \\
& d^5_{1,0,21}: &F_{\alpha_2\bar{\alpha}_3}\left(D_{\bar{\alpha}_1}F_{\alpha_3\bar{\alpha}_4}\right)\left(D_{\alpha_1}F_{\alpha_4\bar{\alpha}_2}\right) \nonumber \\
& d^5_{1,0,22}: &\left(D_{\alpha_1}F_{\alpha_2\bar{\alpha}_3}\right)F_{\alpha_3\bar{\alpha}_4}\left(D_{\bar{\alpha}_1}F_{\alpha_4\bar{\alpha}_2}\right) \nonumber \\
& d^5_{1,0,23}: &F_{\alpha_2\bar{\alpha}_3}\left(D_{\alpha_1}F_{\alpha_4\bar{\alpha}_2}\right)\left(D_{\bar{\alpha}_1}F_{\alpha_3\bar{\alpha}_4}\right) \nonumber \\
\end{eqnarray}
Structure 1: Chains: 1 Loops: 2
\begin{eqnarray}
& d^5_{1,1,0}: &F_{\alpha_2\bar{\alpha}_1}\left(D_{\bar{\alpha}_2}D_{\alpha_1}F_{\alpha_3\bar{\alpha}_4}\right)F_{\alpha_4\bar{\alpha}_3} \nonumber \\
& d^5_{1,1,1}: &F_{\alpha_2\bar{\alpha}_1}F_{\alpha_3\bar{\alpha}_4}\left(D_{\bar{\alpha}_2}D_{\alpha_1}F_{\alpha_4\bar{\alpha}_3}\right) \nonumber \\
& d^5_{1,1,8}: &\left(D_{\alpha_2}D_{\bar{\alpha}_1}F_{\alpha_3\bar{\alpha}_4}\right)F_{\alpha_1\bar{\alpha}_2}F_{\alpha_4\bar{\alpha}_3} \nonumber \\
\end{eqnarray}

\subsubsection*{Superstructure 2: \#derivatives: 4, \#$F$s: 2}

Structure 0: Chains: 0 0 Loops: 2
\begin{eqnarray}
& d^5_{2,0,52}: &\left(D_{\alpha_2}D_{\alpha_1}F_{\alpha_3\bar{\alpha}_4}\right)\left(D_{\bar{\alpha}_2}D_{\bar{\alpha}_1}F_{\alpha_4\bar{\alpha}_3}\right) \nonumber \\
& d^5_{2,0,53}: &\left(D_{\bar{\alpha}_2}D_{\bar{\alpha}_1}F_{\alpha_3\bar{\alpha}_4}\right)\left(D_{\alpha_2}D_{\alpha_1}F_{\alpha_4\bar{\alpha}_3}\right) \nonumber \\
& d^5_{2,0,54}: &\left(D_{\alpha_2}D_{\bar{\alpha}_1}F_{\alpha_3\bar{\alpha}_4}\right)\left(D_{\bar{\alpha}_2}D_{\alpha_1}F_{\alpha_4\bar{\alpha}_3}\right) \nonumber \\
\end{eqnarray}

\end{document}